\documentclass[aps,pra,twocolumn,showpacs,amsmath,amssymb,floatfix,superscriptaddress,notitlepage]{revtex4-2}

\usepackage{color}
\usepackage{bbm}
\usepackage{graphicx}
\usepackage{dcolumn}
\usepackage[utf8]{inputenc}

\usepackage{graphicx}
\usepackage{epstopdf}
\usepackage{amsthm}
\usepackage{amsmath}
\usepackage{empheq}
\usepackage{bbm}
\usepackage{braket}
\usepackage{amssymb}
\usepackage[usenames,dvipsnames]{xcolor}
\definecolor{pythonblue}{RGB}{59, 117, 175}
\usepackage{cases}
\usepackage{latexsym}

\usepackage[colorlinks=true,citecolor=pythonblue,linkcolor=pythonblue,urlcolor=pythonblue]{hyperref}
\usepackage[title]{appendix}
\usepackage{mathrsfs}

\DeclareMathOperator{\tr}{tr}
\DeclareMathOperator{\Tr}{Tr}

\begin{document}

\title{Energy dynamics in a class of local random matrix Hamiltonians}

\author{Klée Pollock}
\affiliation{Department of Physics and Astronomy, Iowa State University, Ames, Iowa 50011, USA}
\email{kleep@iastate.edu}

\author{Jonathan D. Kroth}
\affiliation{Department of Physics and Astronomy, Iowa State University, Ames, Iowa 50011, USA}

\author{Nathan Pagliaroli}
\affiliation{Department of Mathematics, Western University, London ON N6A 3K7, CA}
\email{npagliar@uwo.ca}

\author{Thomas Iadecola}
\affiliation{Department of Physics and Astronomy, Iowa State University, Ames, Iowa 50011, USA}
\affiliation{Ames National Laboratory, Ames, Iowa 50011, USA}

\author{Jonathon Riddell}
\affiliation{School of Physics and Astronomy, University of Birmingham, Edgbaston, Birmingham, B15 2TT, UK}
\email{j.p.riddell@bham.ac.uk}

\begin{abstract}
Random matrix theory yields valuable insights into the universal features of quantum many-body chaotic systems. Although all-to-all interactions are traditionally studied, many interesting dynamical questions, such as transport of a conserved density, require a notion of spatially local interactions. We study the transport of the 
energy, the most basic conserved density, in few-body and 1D chains of nearest-neighbor random matrix terms that square to one, which we dub the ``Haar-Ising" local random matrix model. In the few-body but large local Hilbert space dimension case, we develop a mapping for the energy dynamics to a single-particle hopping picture. This allows for the computation of the energy density autocorrelators and an out-of-time-ordered correlator of the energy density. In the 1D chain, we numerically study the energy transport for a small local Hilbert space dimension. Throughout, we discuss the density of states and touch upon the relation to free probability theory.

\end{abstract}
	
\maketitle

\section{Introduction}

Interacting many-body quantum systems sit at the core of many areas of contemporary physics. At the intersection of condensed matter theory, quantum information theory, and quantum chaos is the broad question of when and how such systems dynamically approach thermal equilibrium \cite{Short_2011,reimann_foundation_2008}. As far as the approach towards thermal equilibrium is concerned, independently of certain details such as interacting spins or interacting electrons, one expects the low-frequency (long-time) physics to be describable by classical hydrodynamic equations that depend only on symmetries and conservation laws. What is particularly intriguing is how a strongly interacting system, which does not admit well-defined quasi-particles, can undergo a hydrodynamic process like diffusion, which usually arises as a coarse-grained description of particles executing a random walk.

Since the microscopic analytical treatment of specific strongly interacting systems (e.g. Hubbard model, QCD, quantum magnets, etc.) is essentially nonexistent, one can appeal to various phenomenological calculations to achieve analytical control. These include mean-field theory, large $N$ limits, conformal symmetry, or, for questions of ``generic" or ``statistical" properties, random matrix theory (RMT). Since Wigner and his successors' work on the statistics of the energy levels of large nuclei \cite{wigner1967random}, Hamiltonian random matrix theory has informed our understanding of chaos in the quantum mechanical setting \cite{bohigas_characterization_1984}. This is particularly true in systems where no semi-classical limit is available \cite{d'alessio_from_2016,cotler_chaos_2017}.

Standard random matrix theory phenomenology treats a zero-dimensional system in the form of all-to-all interactions. Various aspects of locality have, however, been incorporated in the literature. These include various types of banded random matrices \cite{mirlin_1996} whose matrix elements decay away from the main diagonal, Wegner's $n$-orbital model describing disordered hopping in a large $n$ limit \cite{wegner_1979}, or Rosenzweig-Porter models whose diagonal elements are stronger than off-diagonal, but which still maintain all-to-all interaction  \cite{kravtsov_2015}. Local interactions also give rise to sparse Hamiltonians, which have been treated \cite{rodgers_1998}, along with Sachdev-Ye-Kitaev (SYK) models which can also be viewed as sparse random matrices and are discussed later in this section. Certain manifestations of interactions (through correlations) have also been introduced in the RMT literature \cite{shukla_2005}.

Many of these models succeed in describing universal physics at the Heisenberg timescale (inverse of the typical many-body level-spacing) of various interacting chaotic systems or disordered non-interacting ones (e.g., level statistics in the localized and delocalized phases and at the critical point of the Anderson localization transition). Our primary motivations are somewhat different. We are interested primarily in dynamics of a generic Hamiltonian system subject to geometric locality of interaction, for which one can treat questions of hydrodynamics of a local conserved density. Such questions also involve ``early-time" dynamics, e.g. $t\sim O(L^\alpha)$ in a system of $L$ local subsystems, as opposed to the Heisenberg time-scale $t\sim e^{O(L)}$ that is relevant for level statistics.

On the other hand, the properties of generic local and unitary dynamics without energy conservation have been well understood by random quantum circuit calculations \cite{fisher_random_2023}. One can build in simple conservation laws to the dynamics and derive emergent hydrodynamics of those densities \cite{khemani_operator_2018}, even for the circuit-to-circuit fluctuations \cite{mcculloch_full_2023}. Operator spreading in these models without any conservation laws can also be understood as a hydrodynamic process \cite{von_keyserlingk_operator_2018}. In these random circuit calculations, one can consider models with and without (discrete) time translation invariance \cite{chan_many-body_2022}, but in either case the price one pays is a lack of strict energy conservation. 

It is interesting to ask if any of these tools can be extended to treat generic local energy conserving dynamics on the lattice. Then, one could try to prove that the presence of the most fundamental local conservation law, energy, leads to a universal emergent hydrodynamic description. Energy transport is also interesting because the globally conserved charge generates the dynamics itself. Unlike the case of spin transport, for energy transport the local charge density operators \emph{must} fail to commute (otherwise the system would be non-interacting) making analytical treatment more challenging. One line of work treats the question of energy transport for the Sachdev-Ye-Kitaev (SYK) model \cite{sachdev_gapless_93,kitaev_simple_2015} and indeed diffusive energy transport has been established in 1D generalizations thereof \cite{gu_local_2017,zanoci_energy_2022}. Here we are interested in such calculations for a random matrix generalization of a disordered locally interacting 1D quantum spin chain. For a pair of random matrices interacting in a strictly zero-dimensional manner, the energy dynamics were derived when the total density of states has a square-root dependence on energy at the edge of the spectrum (as is common in random matrix theories) \cite{bellitti_2019}. Our work should be seen as complementary to that one, which will be referenced throughout.
 
There is also the simpler question of the equilibrium physics of such disordered systems with local interactions. This is controlled by the partition function and therefore the density of states. The density of states of the so-called double-scaled SYK model was computed exactly \cite{berkooz_towards_2019,verlinde_double-scaled_2024} and tunes between a Gaussian and semicircular distribution as a function of the number of bodies in the SYK interaction (when properly scaled). A transition from semicircle to Gaussian in quantum spin glasses on varying graphs was also identified \cite{erdos_phase_2014}.  Some works have begun to address the question of locality for the density of states of a local random matrix in the large local Hilbert-space dimension limit \cite{morampudi_many-body_2019,collins_spectrum_2023,speicher_mixtures_2016}.

\subsection{Local Haar-Ising random matrix model}

\begin{figure}[t]
    \centering
    \setlength{\abovecaptionskip}{-5pt}
    \includegraphics[width=0.75\columnwidth]{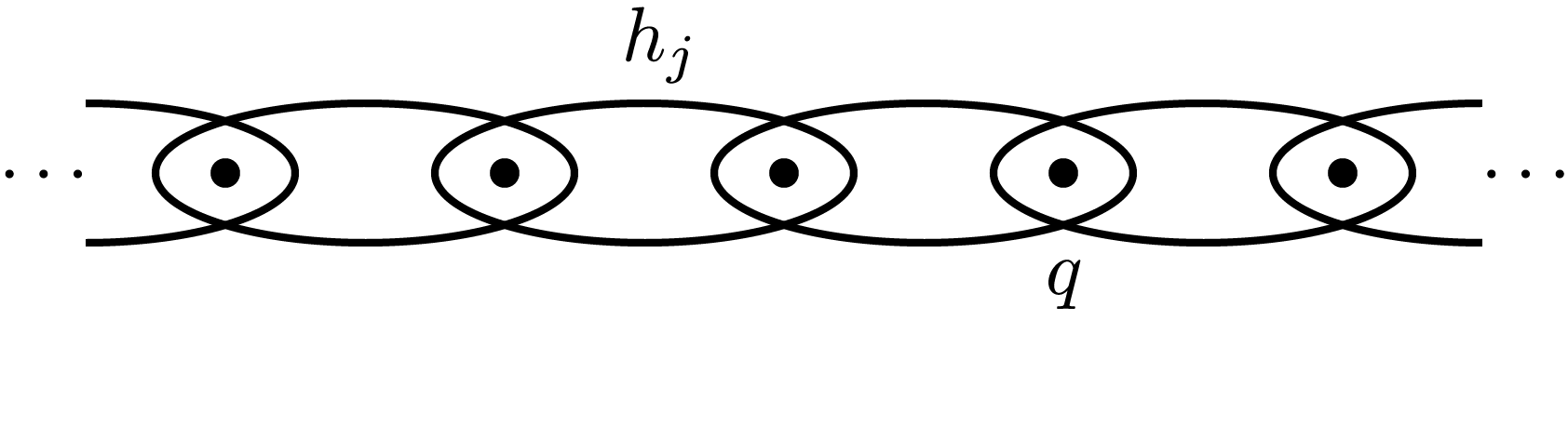}
    \caption{A local nearest-neighbor random matrix chain. Each site is a $q$-dimensional Hilbert space and the bonds are the random matrix interactions.}
    \label{fig:full_model}
\end{figure}

In this paper we study the following ensemble of locally interacting random matrix Hamiltonians. The Hamiltonian acts on a 1D lattice of local $q$-dimensional systems as
\begin{equation}\label{eq:full_model}
    H = \sum_j h_j,
\end{equation}
where $h_j$ is a nearest-neighbor (NN) spin-spin like interaction term coupling sites $j,j+1$ (see Figure \ref{fig:full_model}). We also refer to the interaction operators $h_j$ as ``bonds". Each local term is realized as $h_j = U_j \Lambda U^\dagger_j \otimes I_{\overline{j}}$ for which $\Lambda^2=1$, $\text{tr}(\Lambda)=0$, and $U_j$ are independent Haar random unitary matrices. The matrices $U_j$ and $\Lambda$ are $q^2\times q^2$, reflecting the NN locality of interaction, and $I_{\overline{j}}$ is an identity matrix acting on all sites besides $j,j+1$. We refer to $U_j \Lambda U^\dagger_j$ as Haar-Ising (HI) random matrices since their eigenvectors are Haar random and their eigenvalues are $\pm 1$ in equal number. We will consider both periodic (PBC) and open boundary conditions (OBC). A translation invariant version with $U_j = U$ could also be considered, though we focus on the disordered case in this paper. In the free probability literature, a self-adjoint random variable with this spectrum is referred to as a symmetric Bernoulli variable \cite{nica_lectures_2006} and such matrices were also studied in \cite{bellitti_2019} under the name ``binary" random matrices.

The following features of the model Eq.~\eqref{eq:full_model} make it an interesting platform for studying generic out-of-equilibrium Hamiltonian dynamics. First, the local terms generalize Ising or spin 1/2 Heisenberg interactions of the form $\vec{S}_j\cdot \vec{S}_{j+1}$. However, we note that large $q$ does not correspond to large spin $S$. Secondly, the local terms have individually bounded eigenvalue spectra of $\pm1$ even in a large $q$ limit, and so the Hamiltonian strictly obeys Lieb-Robinson bounds \cite{lieb_finite_1972} independently of $q$ (for each arbitrarily large but finite $q$). Additionally, the maximum eigenvalue of the full Hamiltonian is bounded by $L$ independently of $q$, so there is no subtlety related to rescaling the energy with $q$ in order to make the system thermodynamically well-defined at large $q$.

Thirdly, unlike a chain of local terms drawn from the Wigner-Dyson ensembles (GUE, GOE or GSE), the terms we study are not individually quantum chaotic since they have massively degenerate eigenvalue spectra. This is interesting from a many-body quantum chaos point of view, since the putative chaoticity of the model would only arise through the interplay of the terms. One can imagine that the dynamics of the spectral form factor (SFF) \cite{mehta_random_2004} would differ from a theory with locally chaotic terms, where (at least in discrete time) there is a transition from the early time behavior $t^L$ characteristic of $L$ non-interacting chaotic systems, to the universal late-time ramp $t$ \cite{chan_spectral_2018}. Considering two interacting terms in particular, perhaps in the GUE case there is an early time $t^2$ behavior of the connected SFF, whereas for the HI case there would be no such regime. We reserve a study of chaos for future work. We also note that quantum chaos  through the perspective of the eigenstate thermalization hypothesis (ETH) was studied in the situation of local RMT Hamiltonians \cite{test_sugimoto_2021,eigenstate_sugimoto_2022}.

\subsection{Energy dynamics}

In this study, we are principally interested in the dynamics of the local energy density operators $h_j$. In particular, we focus on random matrix averaged, infinite temperature, two point functions of the local density, which we denote
\begin{equation}
    C_{ij}(t) = \braket{h_i(t)h_j(0)}.
\end{equation}
Throughout, we write $A(t)=e^{-iHt}Ae^{iHt}$ for the Heisenberg evolution of an operator $A$. We employ the notation $\tr (A) :=  \frac{1}{N}\Tr (A)$, when $A$ is $N\times N$, to refer to a normalized trace. Lastly, we use $\langle \cdot \rangle := \mathbb{E}[\tr(\cdot)]$ where the random matrix average $\mathbb{E}[\cdot]$ refers to averaging over the random matrix terms present in the expression. 

These two-point functions are of interest because $C_{ij}(t)$ is the infinite-temperature linear response coefficient of bond $i$ to a weakly out-of-equilibrium initial configuration 
\begin{equation}
    \rho_j(\mu) = e^{\mu h_j} / \Tr(e^{\mu h_j})
\end{equation}
corresponding to an energy $\mu$ on bond $j$ and zero (on average) elsewhere. One can then construct the linear response to an arbitrary initial configuration of the energy from the $C_{ij}(t)$. We remark that for the HI local terms that satisfy $h_j^2=1$, the response becomes exactly
\begin{equation}
    \braket{h_i(t)\rho_j(\mu)} = \sinh(\mu) C_{ij}(t),
\end{equation}
implying that $C_{ij}(t)$ is not just a linear response coefficient, but the total infinite temperature response of the system to an initial condition of unit energy (after dividing through by $\sinh (\mu)$) present on one of the bonds for the model we study.

For a generic interacting 1D system with no other conservation laws beyond energy, we would expect the high-temperature energy transport to be diffusive and $C_{ij}(t)\rightarrow t^{-1/2}$ as $t \rightarrow \infty$ (after first taking $L\rightarrow \infty$) \cite{Bertini2021}. This is to be contrasted with an integrable system that has extensively many conservation laws and where we would expect $C_{ij}(t)\rightarrow t^{-1}$ corresponding to ballistic motion of particles carrying the energy. Interestingly, the energy transport in the integrable XXZ chain remains ballistic as the anisotropy is tuned, consistent with integrability, even when spin transport becomes diffusive \cite{gopalakrishnan_anamalous_2023}.

Anomalous energy transport is also possible in non-integrable systems; for example, in the presence of kinetic constraints such as PXP dynamics, Kardar-Parisi-Zhang superdiffusion was found \cite{ljubotina_superdiffusive_23}. If the system is disordered, one might anticipate localization phenomena, and, for example, a sufficiently disordered XYZ chain exhibits sub-diffusion \cite{schulz_energy_18} of energy at high temperature. However, despite our $q=2$ Haar-Ising random matrix spin chain being in some sense maximally disordered, we do not find any evidence of energy localization, and instead find certain numerical signatures of diffusion. The lack of localization is consistent with the level spacing statistics being that of a GUE and not Poisson (which would correspond to many-body localization \cite{abanin_colloquium_19}).

Beyond two-point functions relevant to transport of energy, we also investigate out of time ordered correlators (OTOCs) 
\begin{equation}
    O(t) = \frac{1}{2}\braket{[h(t),h(0)]^\dagger [h(t),h(0)]},
\end{equation}
where we fix the location of $h = h_j$ for both terms. OTOCs feature rich markers for the dynamics present in quantum many-body systems. Initially proposed as a probe of quantum chaos similar to Poisson brackets \cite{Maldacena_2016}, OTOCs capture the locality of the model \cite{Lin_2018,Riddell2019}, quantum Lyapunov exponents \cite{Khemani_2018,Xu_2019} and wave-front dynamics \cite{Riddell2023}. Additionally, the late time dynamics carry some information about quantum chaos \cite{Fortes2019,Fortes_2020,Garc_a_Mata_2023,Riddell2020}. In this work we use OTOCs primarily as a probe of quantum scrambling, and as a comparison to known random matrix theory results \cite{vijay_finite-temperature_2018}. 

\subsection{Summary and organization of results}

In this paper, we survey various limits of the model Eq.~\eqref{eq:full_model}, focusing on the energy dynamics as well as the density of states. 

In Sec.~\ref{sec:z2} we study two Haar-Ising terms overlapping a single site on a three site lattice. We show that infinite-temperature equal-time correlators of the local terms are either one or zero in the large $q$ limit, which also proves that the two terms are asymptotically free in the language of free probability \cite{mingo_free_2017}, despite their eigenvectors not being related by a full Haar random unitary. The proof also follows from a recent result in free probability \cite{collins_spectrum_2023}. With this, we establish a mapping to a free particle hopping on a 1D lattice and use it to analytically compute the density of states, energy-energy correlators, and an OTOC at infinite temperature. We also calculate the energy-energy correlator at finite temperature.

In Sec.~\ref{sec:higher_z} we generalize this mapping to treat $z$ overlapping Haar-Ising terms in a situation where none commute. We map the dynamics to a particle hopping on the Bethe lattice with coordination number $z$ and then compute the density of states and energy-energy correlators. We will see that in all of these cases $z\geq 2$, the dynamics of the local energy density does not follow a diffusive form. This is due to the few-body nature of the Hamiltonian, which is local in the sense that it is more sparse than a full random matrix, but not geometrically local as in the full extended chain. 

Ultimately, we would like to treat the energy transport analytically in this full extended 1D HI chain. Since the introduction of true locality (in the sense of some terms commuting and some not) makes the analysis more difficult, we initiate the study by numerically calculating the energy transport for a $q=2$ chain of up to $L=22$ spins. We scrutinize the emergence of energy diffusion in the system by comparing against the solution to the diffusion equation on the same finite lattice, and find that while the autocorrelator quickly converges to that form, the spatially distant two-point functions are slower to converge.

In Sec.~\ref{sec:free} we also address the simpler question of the density of states of various few-body and extended HI chains and make the relation of our previous analysis to free probability theory more explicit. We numerically identify some situations where standard techniques from free probability can be naively applied to compute the density of states, and some where they cannot. 

We conclude in Sec.~\ref{sec:conc} with some comments about how we may generalize the single-particle mapping, including numerical evidence that only the reducible words survive a large $q$ limit more generally, as well as some remarks about chaos.

\section{Two-term ``chain" in OBC}\label{sec:z2}

\begin{figure*}
    \centering
    \setlength{\abovecaptionskip}{-15pt}
    \includegraphics[width=2.05\columnwidth]{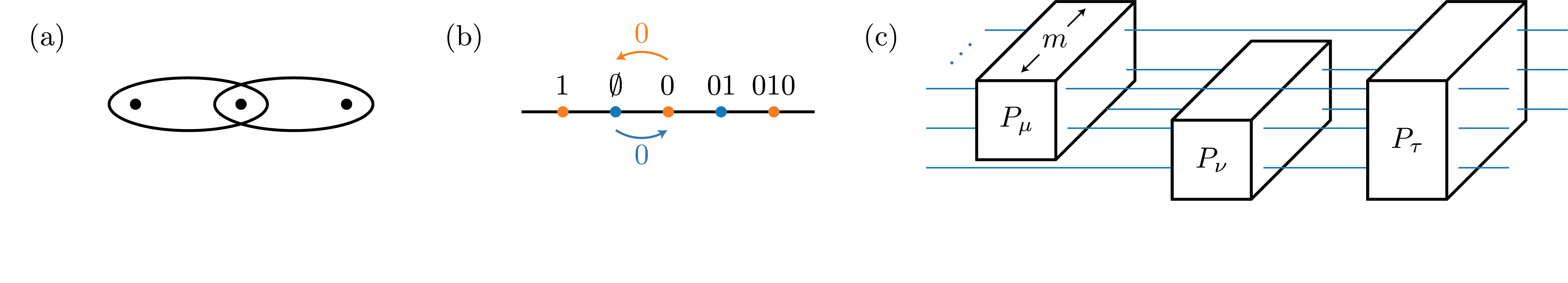}
    \caption{Panel (a) depicts two interacting HI bonds on a three site lattice. Panel (b) is an infinite 1D lattice on which particle trajectories correspond to words of two non-commuting HI terms. The origin corresponds to the empty word $\emptyset$ and each other site corresponds to a fully reduced (after applying $h_i^2=1$) irreducible word. Panel (c) is a tensor network diagram used to evaluate zero-time correlators $\braket{(h_0h_1)^m}$ of the terms in model (a), where one averages over $m$ replicas of the three site system. After averaging, one is left with a trace over certain permutation matrices $P_\sigma$, see Eq.~\eqref{eq:trace} in the text.}
    \label{fig:two_terms}
\end{figure*}

In this section, we show how a large $q$ random matrix average allows a simple calculation of any correlation function composed of only energy density operators for a pair of interacting HI random matrices.  The Hamiltonian is
\begin{equation}\label{eq:model}
    H = h_0 + h_1,
\end{equation}
with $h_i$ being $q^3\times q^3$ independently drawn random matrices that act locally on a three site lattice as in Fig.~\ref{fig:two_terms}(a). To enforce the locality and HI property, the matrices are realized as
\begin{equation}
    h_0 = U_0\Lambda U_0^{\dagger} \otimes I, \quad h_1 =  I \otimes U_1\Lambda U_1^{\dagger},
\end{equation}
with $U_j$ and $\Lambda$ defined as in Eq.~\eqref{eq:full_model}. In this case, $I$ is a $q\times q$ identity. We remark that in Ref.~\cite{bellitti_2019} the energy-energy correlator in a strictly zero-dimensional version of this model (both terms acting on the same site) was considered only numerically, here we derive an analytical result for this quantity.

We first show that ``locality", in the sense of two energy bonds which only partially overlap on a three site lattice, is not analytically prohibitive and leads to the same combinatorics in the large $q$ limit as two HI random matrices acting on the same site. We then show how the combinatorics of any single-trace average of two non-commuting HI random matrix terms can be described in terms of particles hopping on a 1D lattice. This gives way to a simple calculation for the density of states, the local energy-energy OTOCs of the local energy density. In the next section, Sec.~\ref{sec:higher_z}, we discuss generalizing this analysis to $z$ mutually non-commuting HI terms.

\subsection{Alternating correlators vanish}

As discussed in the following sections, we are interested in single-trace correlation functions of the local terms that look like, for example,
\begin{equation}
    \braket{h_0h_0h_1h_0h_1 h_0}.
\end{equation}
Given such a word of the local terms, we can apply the algebraic relations $h_i^2=1$ to fully reduce the word.  If it cannot be reduced to the identity operator, then we call it irreducible. The string of indices of (fully reduced) irreducible words must be of one of the following forms: $(01)^m$,~$(10)^m$,~$1(01)^m$,~$0(10)^m$. Since we are interested in traces of the random matrix terms, the first two strings are equivalent, while the latter two can be further reduced to $0$ and $1$, respectively, due to cyclicity of the trace. Those latter two vanish since $\braket{h_i}=0$ by construction. Regarding the first type, we assert the non-trivial claim that the alternating words $(01)^m$ also vanish in the large $q$ limit, i.e.
\begin{equation}
    \braket{(h_0h_1)^m} \rightarrow 0\quad \text{as}\quad q\rightarrow \infty. 
\end{equation}
This is proven by invoking the Weingarten calculus for averaging over the unitary group, which we find instructive to work out in the remainder of this section. The result also, however, follows from recent work in free probability theory \cite{collins_spectrum_2023,charlesworth_matrix_2021}.

The formulation of the Weingarten calculus for large $q$ derived in \cite{fava_designs_2023} is very useful for our purposes. Let $h = U\Lambda U^{\dagger}$ be a $q^2\times q^2$ Hermitian matrix with Haar eigenvectors. We momentarily relax the assumption that $\Lambda^2=1$. The average $m$-fold operator is
\begin{equation}
    \mathbb{E}~h^{\otimes m} = \mathbb{E}_{U}~ U^{\otimes m} \Lambda^{\otimes m}U^{\dagger \otimes m}
\end{equation}
and can be viewed as the action of the $m$-fold ``Haar channel" on $\Lambda^{\otimes m}$. In the large $q$ limit, the Weingarten calculus tells us  that we obtain the linear combination of permutations
\begin{equation}\label{eq:weingarten}
    \mathbb{E}~h^{\otimes m} \rightarrow \sum_{\mu \in S_m} q^{-2|\mu|} \kappa_\mu(\Lambda^m) P_\mu \quad \text{as}\quad q \rightarrow \infty.
\end{equation}
Here, $S_m$ is the permutation group of order $m$ and $P_\mu$ is the permutation matrix corresponding to $\mu$ that permutes $m$ copies of (in our case) the $q^2$ dimensional Hilbert space that $U$ acts upon. The ``length" $|\mu|$ of a permutation $\mu$ is the minimal number of transpositions needed to represent $\mu$. The length is also equal to $|\mu| = m -|\text{Orb}(\mu)|$ where  $|\text{Orb}(\mu)|$ is the number of orbits of the permutation.

The coefficients $\kappa_\mu(\Lambda)$ are the product of the free cumulants of $\Lambda$ according to the partition of $m$ induced by the cycle structure of $\mu$. More precisely, if the orbits are of lengths $\ell_{1},\ell_{2},...,\ell_{p}$, then the free cumulant is defined as 
\begin{equation*}
    \kappa_{\mu} (\Lambda):= \prod_{i=1}^{p}\kappa_{\ell_{i}}(\Lambda),
\end{equation*}
where $\kappa_n(\Lambda)$ is the usual $n$-th free cumulant studied in free probability theory \cite{nica_lectures_2006}. For example, if $\mu=(1)(234)(5 6)$  in cycle notation, then $\kappa_\mu(\Lambda)=\kappa_1(\Lambda)\kappa_3(\Lambda)\kappa_2(\Lambda)$. Table~\ref{tab:free} states a few examples of the $\kappa_\mu$, but we refer the reader to \cite{fava_designs_2023} for an accessible exposition of the topic.

\begin{table}[h]

\renewcommand{\arraystretch}{1.5}
\setlength{\tabcolsep}{12pt}
\begin{tabular}{|c|c|}
\hline
$\mu$ & $\kappa_\mu(\Lambda)$ \\
\hline
$(1)(2)(3)(4)$ & $\tr[\Lambda] ^4 =0$ \\
$(1234)$ & $\tr[\Lambda^4]-2\tr[\Lambda^2]^2 -\tr[\Lambda]^4 = -1$ \\
$(12)(34)$ & $(\tr[\Lambda^2]-\tr[\Lambda]^2)^2 = 1$ \\
\hline
\end{tabular}
\caption{Some examples of the free cumulants $\kappa_\mu$ and their values for the $\Lambda$ under consideration. \label{tab:free}}
\end{table}

Armed with the Weingarten calculus, we now appeal to tensor network diagrams to better understand the alternating correlators. It will be useful to first express
\begin{equation}
    \braket{(h_0h_1)^m} = \braket{h_0^{\otimes m} h_1^{\otimes m} P_\tau}
\end{equation}
where $\tau=(123\dots m)$ is the full forward cycle in $S_m$ and $P_\tau$ cyclically permutes all three physical legs simultaneously between the $m$ copies. Applying Eq.~\eqref{eq:weingarten} to each independent term, this evaluates to
\begin{equation}
    \braket{(h_0 h_1)^m} \rightarrow \sum_{\mu \nu} \frac{\kappa_\mu(\Lambda) \kappa_\nu(\Lambda)}{q^{2|\mu|+2|\nu|}} \tr[(P_\mu \otimes I)(I \otimes P_\nu)P_\tau]
\end{equation}
where $I$ is a $q\times q$ identity and  $P_\sigma$ is a permutation matrix which can permute different Hilbert spaces according to $\sigma$, which is context dependent. For clarity, Fig.~\ref{fig:two_terms}(c) shows how the different $P$'s act. This figure shows the operator $(P_\mu \otimes I)(I \otimes P_\nu)P_\tau$. The three blue wires are the three physical sites, and moving back into the page are the $m$ replicas of the whole Hilbert space. The last step is to take the normalized trace of this operator:
\begin{multline}\label{eq:trace}
     \tr[(P_\mu \otimes I)(I \otimes P_\nu)P_\tau] \\
     = q^{-3} q^{|\text{Orb}(\tau \mu)|} q^{|\text{Orb}(\tau \nu)|} q^{|\text{Orb}(\tau \nu \mu)|}.
\end{multline}
Therefore,
\begin{equation}\label{eq:after_avg}
    \braket{(h_0 h_1)^m} \rightarrow \sum_{\mu, \nu \in S_m} \kappa_\mu(\Lambda) \kappa_\nu(\Lambda) q^{-2[g(\mu,\nu) + g(\mu)+ g(\nu)]},
\end{equation}
where 
\begin{equation}\label{eq:genus}
    2g(\sigma) = |\tau \sigma| + |\sigma| - (m-1)
\end{equation}
and $g(\sigma)$ is the ``genus" of a permutation $\sigma$ \cite{cori_counting_2013}. We remark that the right-hand side of Eq.~\eqref{eq:genus} is always even. This follows from the fact that the standard notion of parity for a permutation $\sigma$ coincides with the parity of the number $|\sigma|$. We also define
\begin{equation}\label{eq:joint_nc}
    2g(\mu,\nu) = |\tau\nu\mu| + |\mu| + |\nu| - (m-1),
\end{equation}
where $g(\mu,\nu)$ measures a kind of joint genus of $\mu$ and $\nu$, which is also integer-valued. Note that from the triangle inequality on lengths of permutations \cite{nica_lectures_2006}
\begin{equation}
    |\alpha\beta| + |\alpha| \geq |\beta|
\end{equation}
we can deduce that both $g(\mu,\nu)$ and $g(\sigma)$ are non-negative. It follows that Eq.~\eqref{eq:after_avg} is finite as $q \rightarrow \infty$.

Now, an important fact is that the odd free cumulants of a symmetric Bernoulli variable and therefore the those of the HI random matrices are zero \cite{nica_lectures_2006}. Therefore, if $\mu$ or $\nu$ contain at least one cycle of odd length, then that term will not contribute to the sum Eq.~\eqref{eq:after_avg}. We will show that consequently, there are no non-zero $O(q^0)$ terms (nor any $O(q^{-1})$ terms) in that sum.  First, if $m$ is odd, both $\mu$ and $\nu$ necessarily contain an odd length cycle so the correlator is identically zero. If $m$ is even, however, the following argument shows that $g(\mu,\nu)\geq 1$. Because $\mu$ and $\nu$ contain only even length cycles, we have
\begin{equation}\label{eq:pairing_bound}
    |\mu|\geq m/2\quad \text{and}\quad |\nu|\geq m/2,
\end{equation}
where the bounds are saturated for pairing permutations. Supposing that $|\tau\nu\mu|\geq1$, the bound $g(\mu,\nu)\geq 1$ is trivially obtained. Otherwise if $|\tau\nu\mu| = 0$, then $\tau\nu\mu$ is equal to the identity permutation, or in other words $\tau^{-1}= \nu\mu$. Since $m$ is even, $\tau^{-1}$ has odd parity and so $\mu$ and $\nu$ must have opposite parity. This means that the bounds Eq.~\eqref{eq:pairing_bound} cannot both be completely saturated, and in particular $|\mu|+|\nu| \geq m + 1$, leading to $g(\mu,\nu)\geq 1$. Finally, since $g(\mu)\geq 0$ and $g(\nu)\geq 0$, we have shown that 
\begin{equation}\label{eq:m_2}
    \braket{(h_0h_1)^m} = O(q^{-2})\quad \text{as}\quad q\rightarrow \infty.
\end{equation}



Consider the $m=2$ case. Here, there is only one term that can contribute in the sum Eq.~\eqref{eq:after_avg} which is $\mu=\nu=(12)$ in cycle notation. The free cumulants are both 1 and so, in the large $q$ limit,
\begin{equation}
    \braket{h_0h_1h_0h_1} \rightarrow q^{-2}
\end{equation}
with no prefactor. We show in Fig.~\ref{fig:two_term_numerics}(a) that this scaling with $q$ can be seen numerically already for $q=8,10,12,\dots$ in this $m=2$ case \footnote{For larger $m$, e.g. $m=4$, the realization-to-realization fluctuations appear to be significantly larger (e.g.  $80 \times$ larger) than the average and so the scaling of the true average is difficult to see numerically.}.

In the large $q$ limit, we can therefore restrict our attention to only the completely reduced correlators of $h_i$. While this suffices for our purposes, it is likely that tighter bounds could be obtained. Computer algebra calculations show that when $m=4$, the alternating correlators are $O(q^{-4})$, and when $m=6$, they are $O(q^{-16})$.

There are two corollaries to this result. One is that $h_0$ and $h_1$ are asymptotically free, which we define and discuss later in Sec.~\ref{sec:free}. Another is that the same result holds in the nearest-neighbor extended chain for $h_j$ and $h_{j+1}$; we have shown that alternating correlators of those terms vanish.

\subsection{Density of states}

The density of states can be calculated through the many-body Green's function $G(u) = \braket{(u-H)^{-1}}$, where the density of states at energy $\epsilon$ is obtained by taking the limit $u=\epsilon+i0^+$ \cite{mingo_free_2017}. Sometimes, it can be deduced via knowledge of the moments $\braket{H^n}$. Let us focus directly on these moments. We have a sum over all possible binary strings of $h_0$ and $h_1$ that are length $n$: 
\begin{equation}\label{eq:moment}
    \braket{H^n} = \sum_{b} \braket{h_b},
\end{equation}
where $h_b$ is shorthand for $h_{b_1}h_{b_2}h_{b_3}\dots h_{b_n} $. We can first reduce all words using the algebraic relations $h_i^2=1$ for $i=0,1$. If $b$ can be fully reduced, the corresponding term with be unity. If it cannot, it will vanish in the large $q$ limit according to Eq.~\eqref{eq:m_2}. Therefore in the large $q$ limit, we need to count the fully reducible strings.

We can map the counting of reducible strings to counting trajectories of a particle on an infinite 1D lattice. A general bitstring (reducible or not) maps to a trajectory as follows. Each bit in the string (starting from the right without loss of generality) corresponds to the particle hopping once. Whether the hop is to the right or to the left, however, depends upon the parity of the current site. If the particle sits at an even site (blue lattice points in Fig.~\ref{fig:two_terms}(b)), a $0$ moves it to the right and a $1$ moves it left, but the opposite holds if the particle is at an odd site (orange lattice points in the same figure). This is set up so that reducible strings correspond to walks that return to their starting location, e.g. the string $00$ corresponds to hopping once right and then left. Note that the lattice points are in one-to-one correspondence with fully irreducible words.

Counting the returning walks is well studied in many different contexts. For example, the calculation of the return probability of a random walk in 1D. For more complicated calculations later, it will be both conceptually and practically useful for us to introduce a nearest-neighbor hopping Hamiltonian
\begin{equation}\label{eq:Delta}
    \Delta = \sum_{x\in \mathbb{Z}} \ket{x-1}\bra{x} + \ket{x+1}\bra{x},
\end{equation}
where $\ket{x}$ is a single particle sitting at site $x$. Since $\ket{0}$ is the particle sitting at the origin, we have for example
\begin{equation}
    \Delta^2 \ket{0} = \ket{-2} + 2\ket{0} + \ket{2}.
\end{equation}
One can then see that the number of reducible words/walks of length $n$ is given by the amplitude to return. Therefore we have the correspondence $\braket{H^n}=\braket{0|\Delta^n|0}$. These vanish for odd $n$, and, for $n=2k$, the sequence is given by the central binomial coefficient
\begin{equation}
    \binom{2k}{k}.
\end{equation}
The random matrix many-body density of states must then be equivalent to the single-particle hopping density of states. This is given by the well known formula
\begin{equation}\label{eq:arcsine}
    \mathcal{D}(\epsilon) = \frac{1}{\pi} (4-\epsilon^2)^{-1/2},\quad \epsilon \in (-2,2)
\end{equation}
for the tight-binding density of states \cite{economou_greens_2006}. Note that this correspondence follows since the probability density is compactly supported and is therefore uniquely determined by its moments.

\begin{figure*}[t]
    \centering
    \includegraphics[width=2.05\columnwidth]{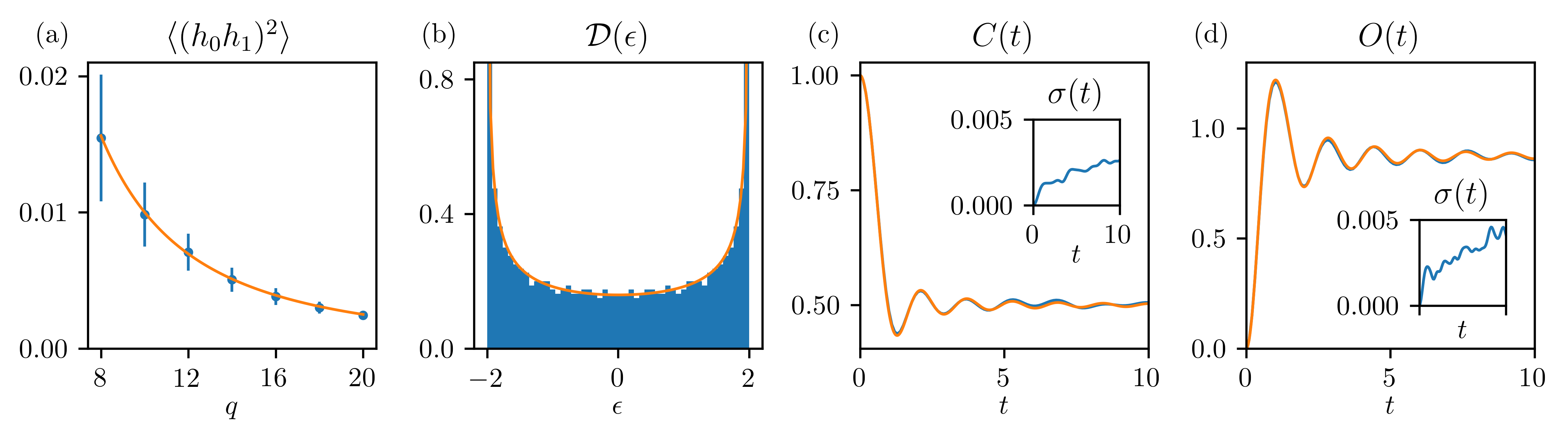}
    \caption{Numerical comparison of the present methods with numerics in blue and analytical theory in orange. This figure concerns single-trace statistics of two independent overlapping HI random matrices. Panel (a) shows the finite-$q$ statistics of an alternating correlator averaged over $100$ realizations of the Hamiltonian and compared against the theoretical average value Eq.~\eqref{eq:m_2}. Panel (b) is a normalized histogram of the eigenvalues of Eq.~\eqref{eq:model} for a \emph{single} $q=10$ realization compared against Eq.~\eqref{eq:arcsine} in orange. Panels (c) and (d) are the $q=10$ dynamics of the energy density correlator $C(t)$ and the energy density OTOC $O(t)$, again for a single realization, and compared against Eq.~\eqref{eq:result} and Eq.\eqref{eq:otoc}, respectively. The insets show one standard deviation $\sigma(t)$ from the average of the corresonding quantity across $100$ realizations of the Hamiltonian.}
    \label{fig:two_term_numerics}
\end{figure*}

We confirm this agrees with a simple exact diagonalization (ED) calculation of a single random realization of Eq.~\eqref{eq:model} for a modest $q=10$. By simply binning the $q^3=1000$ eigenvalues into $50$ bins, one coarse grains the spectrum, and Fig.~\ref{fig:two_term_numerics}(b) shows excellent agreement with the prediction.

\subsection{Local energy correlator}

We will now see that the above correspondence also allows for a simple calculation of the quantity
\begin{equation}\label{eq:autocorr}
    C(t) = \braket{h(t)h(0)}
\end{equation}
where $h$ stands either for $h_0$ or $h_1$ (the correlation functions are identical by reflection symmetry).  Here, $h(t)=e^{-iHt}he^{iHt}$ is the Heisenberg evolution of $h$. 

We first view the problem combinatorially before importing all the tools afforded by the momentum space description of the hopping process. The double series expansion
\begin{equation}\label{eq:double_series}
    C(t) = \sum_{nm} \frac{(-it)^n}{n!} \frac{(it)^m}{m!}  \braket{H^n h H^m h}
\end{equation}
shows that we need to calculate $\braket{H^n h H^m h}$ in the large $q$ limit. Having shown the irreducible words vanish, this is equivalent to counting the number of words whose string of indices is of the form $a0b0$ (with $a$ length $n$ and $b$ length $m$) that are reducible.


Recalling the mapping to the 1D lattice hopping, the string $a0b0$ corresponds to the following process when reading from right to left. Starting at the origin, we first hop right, then undergo an arbitrary walk of length $m$. The particle has undergone the walk dictated by the string $b0$, and it will now be sitting at some lattice point $-m + 1\leq x\leq m + 1$ along the line. The next step must be $0$. Depending upon the parity of $x$ (equivalently whether the corresponding irreducible string ends in a $0$ or a $1$), it will hop left or right. Finally, it will undergo another walk of length $n$ dictated by $a$ which starts at $x\pm1$ and ends at the origin $x=0$.


To describe this process for the hopping dynamics, we define another (Hermitian and unitary) operator
\begin{equation}
    V = \sum_{x\ \text{odd}} \ket{x-1}\bra{x} + \sum_{x\ \text{even}}  \ket{x+1}\bra{x}.
\end{equation}
This hops a particle right or left depending on the parity of the site. The strings of reducible words  $a0b0$ are then counted by the amplitude $\braket{0|\Delta^n V \Delta^m V|0}$, i.e. we have established the correspondence in the large $q$ limit,
\begin{equation}
    \braket{H^n h H^m h} = \braket{0|\Delta^n V \Delta^m V|0}
\end{equation}
and therefore we can re-sum the double series Eq.~\eqref{eq:double_series}:
\begin{equation}\label{eq:effective_V}
    C(t) = \braket{0|V(t)V(0)|0} ,
\end{equation}
where now $V(t) = e^{-i\Delta t} V e^{i\Delta t}$ is the Heisenberg evolved staggered hopping operator $V$ \emph{within the single particle sector}.


Now, we go to momentum space where the calculations become simple. The hopping Hamiltonian $\Delta$ defined in Eq.~\eqref{eq:Delta} describes hopping on an infinite lattice [which is needed for validity of Eq.~\eqref{eq:effective_V}]. We first would like to diagonalize a finite version of this Hamiltonian, write the expressions in momentum space, and then take the limit at the end. If we consider a length $N$ and PBC hopping Hamiltonian, we have the complete orthonormal eigenstates of definite momentum
\begin{equation}
    \ket{k} = \frac{1}{\sqrt{N}}\sum_x e^{ikx} \ket{x},\quad k = \frac{2\pi n}{N} 
\end{equation}
where $n\in \{0,1,2,\dots,N-1\}$.

One more useful fact is the following translation invariance \footnote{This is not exactly trivial; but follows by inserting $I = T_x^\dagger T_x$ where $T_x$ translates by $x$ sites:
\begin{equation*}
    \braket{0|V(t)V(0)|0} = \braket{0|T^\dagger_x T_x V(t) T_x^\dagger T_x  V(0) T_x^\dagger T_x |0}.
\end{equation*}
By using that $\Delta$ is translation invariant, we have
\begin{equation*}
    \braket{0|V(t)V(0)|0}
    = \braket{x|(T_x V T^\dagger_x)(t) (T_x V T^\dagger_x)(0)|x}.
\end{equation*}
If $x$ is even, then $T_x V T^\dagger_x = V$. If $x$ is odd,  then $T_x V T^\dagger_x = V'$ where $V'$ now hops an even site particle to the left which is the opposite action of $V$. By symmetry of the original RMT model under $h_0\leftrightarrow h_1$, we still have the claimed equality for odd $x$.}
\begin{equation}
    \braket{0|V(t)V(0)|0} = \braket{x|V(t)V(0)|x}\quad \forall x.
\end{equation}
This allows us to write $C(t) = \tr_1[V(t)V(0)]$ where $\tr_1[\cdot]$ is an infinite temperature average (normalized trace) over the single-particle sector:
\begin{equation}
    \tr_1[V(t)V(0)] = \lim_{N\rightarrow \infty} \frac{1}{N} \sum_{x} \braket{x|V(t)V(0)|x},
\end{equation}
and we keep in mind that $t\ll N$ as $N\rightarrow \infty$ on the right hand side of this equation. Later, we will see that putting the random matrix model at finite temperature corresponds to an average over the single-particle sector at the same finite temperature. Therefore, the response of a bond to an energy impulse in the two-term HI theory maps to the response of a 1D free particle to a staggered hopping impulse.

The infinite temperature average can also be done in momentum space:
\begin{equation}
    C(t) = \lim_{N\rightarrow \infty} \frac{1}{N} \sum_{k} \braket{k|V(t)V(0)|k},
\end{equation}
and inserting resolutions of identity in the momentum basis leads to
\begin{equation}
    C(t) = \lim_{N\rightarrow \infty} \frac{1}{N}\sum_{kp}
    \braket{k|e^{-i\Delta t}|k} |\braket{k|V|p}|^2 \braket{p|e^{i\Delta t}|p}.
\end{equation}
Here, we have used the fact that the single particle propagator is diagonal in momentum space
\begin{equation}
    \braket{k|e^{-i\Delta t}|p} =  e^{-i\epsilon_k t} \delta_{k,p},
\end{equation}
where $\epsilon_k=2\cos k$ is the dispersion relation. The staggered hopping matrix element is also readily evaluated to
\begin{equation}
    \braket{k|V|p} =  \cos(k) \delta_{k,p} +  i\sin(k) \delta_{k, p+\pi}.
\end{equation}
Since the cross terms vanish upon squaring, we obtain the simple expression
\begin{equation}
    C(t) = \int_{0}^{2\pi} \frac{dk}{2\pi} ( \cos^2(k)+ \sin^2(k) e^{-2i\epsilon_kt} ).
\end{equation}
Above, we have taken the limit of $N\rightarrow\infty$ and converted the sum into an integral. The first term is a constant that evaluates to $1/2$; this is the equilibrium value approached as $t\rightarrow \infty$ and the second term dephases. The second term is a standard integral and we find
\begin{equation}\label{eq:result}
    C(t) = \frac{1}{2} + \frac{J_1(4t)}{4t}
\end{equation}
where $J_1$ is a Bessel function. The amplitude of the late-time decay towards equilibrium is therefore the power law $t^{-3/2}$. This is slower than the Gaussian case $t^{-3}$ \cite{bellitti_2019}, which is physically reasonable since, unlike two GUE's interacting, the individual terms are not chaotic and therefore the dynamics is slower.

The fact that $C(t)$ approaches $1/2$ up to some power-law decay is a proof that, starting from a far out-of-equilibrium initial condition, the (average) system comes to thermal equilibrium within a ``finite" time, i.e. a time-scale not growing with $q$. The infinite $q$ form of Eq.~\eqref{eq:result}
matches well the numerical result for a single $q=10$ realization in Fig.~\ref{fig:two_term_numerics}(c). The realization-to-realization fluctuations, as measured by one standard deviation shown in the inset, are remarkably small, showing that empirically, the average $C(t)$ is also typical of a single realization on the relevant timescale.

\subsection{Out of time ordered correlator}

It is interesting to see one more example of the single-particle hopping picture. We consider an OTOC of the local energy density
\begin{equation}
    O(t) = \frac{1}{2}\braket{[h(t),h(0)]^\dagger [h(t),h(0)]},
\end{equation}
where $h$ stands for either $h_0$ or $h_1$ (but the same $h$ throughout the expression). Then, due to $h^2=1$, we find
\begin{equation}
    O(t) = 1 - \braket{h(t)h(0)h(t)h(0)}.
\end{equation}
Combinatorially, this would correspond to counting the number of words of the form $a0b0c0d0$. In terms of the single-particle picture, we simply replace $h$ with $V$ and $H$ with $\Delta$. To evaluate this in momentum space, we write
\begin{multline}
    \braket{0|V(t)V(0)V(t)V(0)|0} \\
    = \lim_{N\rightarrow \infty} \frac{1}{N}\sum_{kq} \braket{k|V(t)V(0)|q}\braket{q|V(t)V(0)|k}.
\end{multline}
The following matrix element between two momenta
\begin{multline}
    \braket{k|V(t)V(0)|q} = \delta_{kq} (\cos^2(k) +\sin^2(k) e^{-2i\epsilon_k t}) \\
    + i \delta_{k,q+\pi}\cos(k)\sin(k)(1-e^{-2i\epsilon_k t})
\end{multline}
leads ultimately to the expression
\begin{equation}\label{eq:otoc}
    O(t) = \frac{7}{8} + \frac{8(3-8t^2)J_1(4t)-48tJ_0(4t)-3tJ_2(8t)}{64t^3}
\end{equation}
for which the late-time behavior of the amplitude is a $t^{-3/2}$ decay towards $O(\infty)=7/8$. Fig.~\ref{fig:two_term_numerics} shows excellent agreement with numerical calculations. Unlike the SYK model, which displays a non-trivial Lyapunov exponent, the early time behavior of Eq.~\eqref{eq:otoc} goes like $t^2$ which is identical to a fully all-to-all model like a GUE independently of which operator is considered \cite{vijay_finite-temperature_2018}. At late times, however, the relaxation towards $O(\infty)$ is $t^{-3/2}$ which is slower than both the Gaussian case of $t^{-3}$ in the presence of energy conservation \cite{bellitti_2019} and even slower still than $t^{-4}$ without energy conservation \cite{vijay_finite-temperature_2018}.
 
\subsection{Finite temperature}

We can generalize the previous calculations to finite temperature. For simplicity, we focus on only the energy density autocorrelator and we define the correlator at finite temperature to be
\begin{equation}
    C_\beta(t) = \frac{\braket{e^{-\beta H} h(t) h(0)}}{\braket{e^{-\beta H}}}.
\end{equation}

Here, we consider the annealed disorder average, since we expect that finite-temperature annealed and quenched averages generally coincide in a large $q$ random matrix theory \cite{cotler_black_2017}. Future work could examine exactly how concentrated the quantity $\text{tr}(e^{-\beta H })$ is around its average for Hamiltonian Eq.~\eqref{eq:model} and therefore bound the difference in the two ways of averaging.

In the large $q$ limit, we obtain, in terms of the single particle hopping picture,
\begin{equation}
    C_\beta(t) = Z_1^{-1}\tr_1[e^{-\beta \Delta} V(t)V(0)]
\end{equation}
and $Z_1= \tr_1[e^{-\beta\Delta}]=I_0(2\beta)$ where $I_0$ is a Bessel function and $Z_1$ the (normalized) partition function in the single-particle sector. This amounts to making the propagator on the left of $V(t)$ pick up a Boltzmann weight for the single particle modes and so we have
\begin{equation}
    C_\beta(t) = \int_{0}^{2\pi} \frac{dk}{2\pi} \frac{e^{-\beta \epsilon_k}}{Z_1} ( \cos^2(k)+ \sin^2(k) e^{-2i\epsilon_kt} ).
\end{equation}
In particular, one can see from the dispersion relation $\epsilon_k = 2\cos k$ that as $\beta\rightarrow\infty$, 
\begin{equation}
    \frac{e^{-\beta \epsilon_k}}{Z_1} \rightarrow \delta(k-\pi).
\end{equation}
This implies that for any fixed $t$, 
\begin{equation}
    C_\beta(t) \rightarrow 1\quad \text{as}\quad \beta\rightarrow \infty,
\end{equation}
i.e. at zero temperature the system does not thermalize at all; the unit injected energy remains on only one bond. This is distinct from the GUE case where it is argued that the correlator would still decay from the initial condition as a power-law with half the exponent of the infinite temperature one when first taking $\beta\rightarrow \infty$ \cite{bellitti_2019}. The above integral can be evaluated to
\begin{equation}
    C_\beta(t) = \frac{1}{I_0(2\beta)}\left[I_2(2\beta) + \frac{I_1(2\beta)}{2\beta} + \frac{J_1(4t-2i\beta)}{4t-2i\beta}\right]
\end{equation}
and so we can also consider first taking $t\rightarrow \infty$ for a fixed $\beta$. In that limit, the amplitude of the decaying part goes as
\begin{equation}
    \bigg|\frac{J_1(4t-2i\beta)}{4t-2i\beta}\bigg| \propto t^{-3/2} \bigg[1- \frac{3}{4}\bigg(\frac{\beta}{2t} \bigg)^2 + \cdots \bigg]
\end{equation}
neglecting an overall prefactor dependent on $\beta$. This is the identical decay to infinite temperature at the leading order, except that  a finite $\beta$ becomes a new timescale in the problem with the ``universal" $t^{-3/2}$ behavior setting in only for $t \gg \beta$.

\section{Arbitrary number of mutually non-commuting terms}\label{sec:higher_z}

\begin{figure}
    \centering
    \includegraphics[width=0.95\columnwidth]{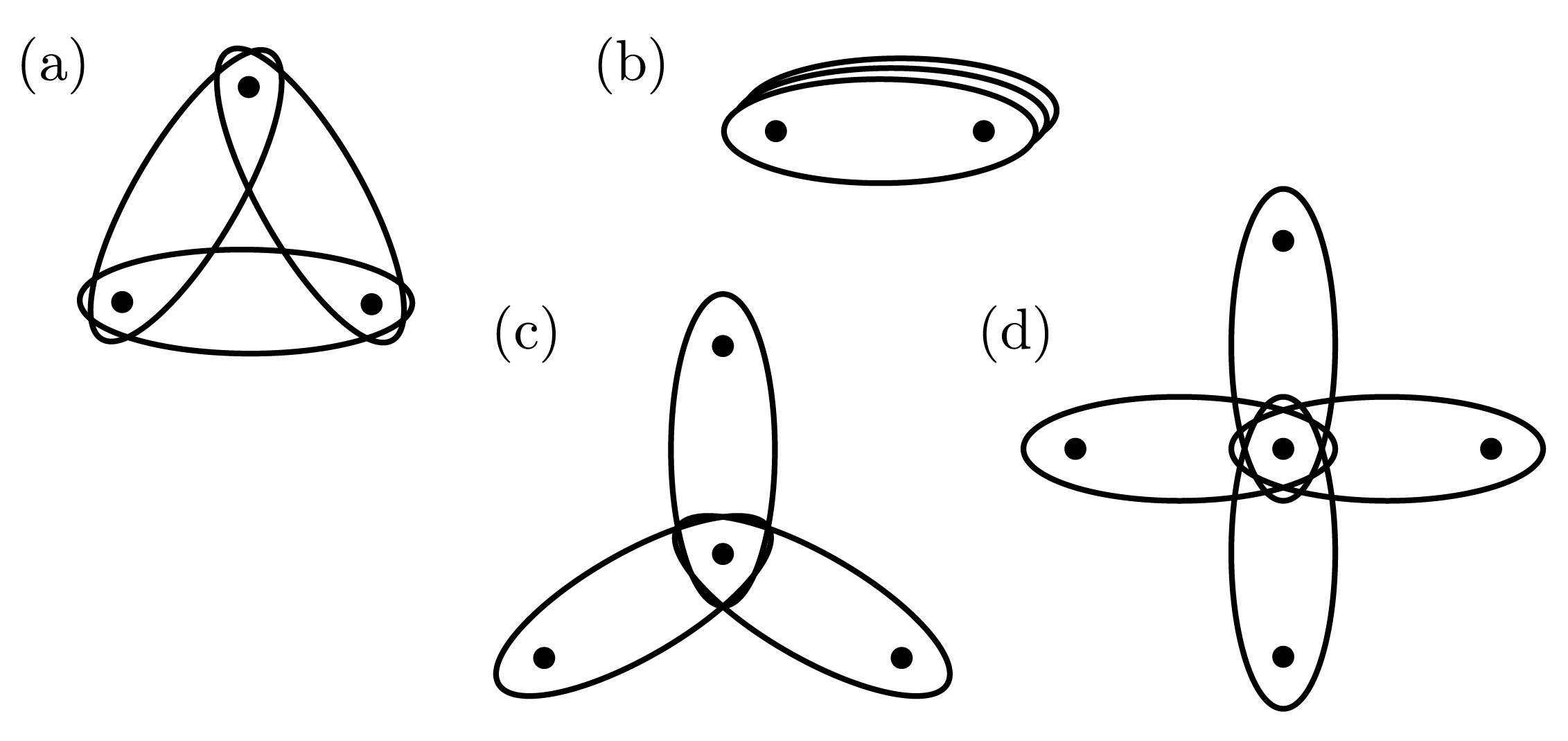}
    \caption{Various few-body scenarios of HI local terms interacting. (a),(b),(c) all map to a single particle hopping on the Bethe lattice with coordination number $z=3$ (for scenario (b) the irreducible words provably vanish). Sketch (a) can be viewed as a short 1D chain in PBC. Sketch (b) is strictly zero dimensional with all $z=3$ terms acting on the same pair of sites. Sketches (c) and (d) can be viewed as parts of a larger lattice of local terms interacting at a site in 2D, and (d) maps to the $z=4$ Bethe lattice.}
    \label{fig:z_sketches}
\end{figure}

One natural extension of the above theory is to add more non-commuting terms. For this to make sense as a ``local" RMT, we could consider three terms in a 1D PBC chain, as in Fig.~\ref{fig:z_sketches}(a) or more bonds all overlapping on a single site as in (c) and (d) of that figure. These are all zero-dimensional systems in the sense that the algebra of the local terms is trivial: all terms are mutually non-commuting (unlike a truly local Hamiltonian). In fact, one can ask simply about multiple HI terms acting on the same pair of sites in a strictly zero-dimensional or all-to-all way, as sketched in Fig.~\ref{fig:z_sketches}(b). While these models are not truly local, their dynamics is still an interesting intermediate question; in particular one may wonder how changing the number, say $z$, of non-commuting HI terms affects the speed of thermalization. Therefore, we are interested in the Hamiltonian
\begin{equation}
    H = \sum_{j=0}^{z-1} h_j,\quad [h_i,h_j] \neq 0\ \forall i\neq j
\end{equation}
where, depending on context, the terms $h_j$ may act on different sites. 

In the previous section, we proved that irreducible words of two terms vanish in the large $q$ limit, when arranged so that they overlap on a single site. This result continues to hold when considering $z$ terms which are mutually non-commuting and arranged with all bonds overlapping on at least a single site, which follows from Refs.~\cite{collins_spectrum_2023,charlesworth_matrix_2021}. This observation gives way to a similar single-particle mapping as developed in Sec.~\ref{sec:z2}, but here the particle hops on the Bethe lattice with coordination number $z$. We study in a bit more detail the particular case of $z=3$ with the terms arranged as in Fig.~\ref{fig:z_sketches} panel (a), confirming that numerically, the single particle hopping on the Bethe lattice gives the empirically correct density of states and energy-energy correlator.

\begin{figure*}[t]
    \centering
    \setlength{\abovecaptionskip}{-5pt}
    \includegraphics[width=2.05\columnwidth]{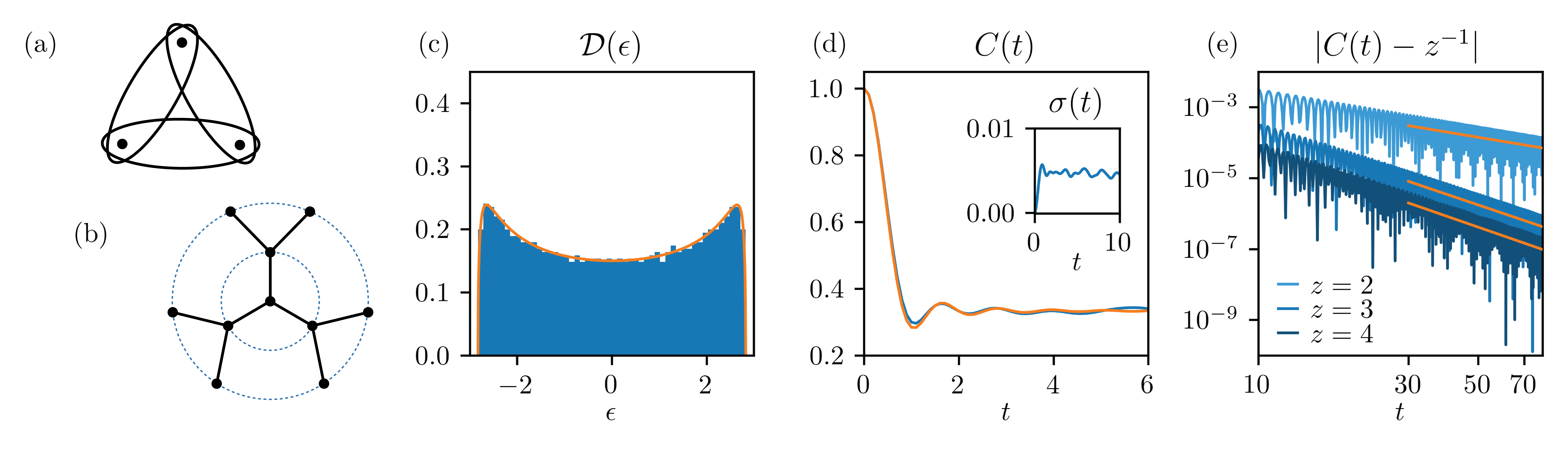}
    \caption{Sketch (a) is an example of ``local" random matrix with pairwise interactions that can be viewed as a 1D ``chain" in PBC. Sketch (b) is a portion of the $z=3$ Bethe lattice. Panel (c) compares the empirical density of states for a single $q=12$ realization of model (a) with the theory, Eq.~\eqref{eq:dos_z}. Panel (d) compares (blue) the numerically calculated local energy autocorrelator of a single $q=6$ realization of model (a) to a certain hopping dynamics on the Bethe lattice, Eq.~\eqref{eq:effective_X} (orange). The latter is also calculated numerically on a finite ``radius" $R=300$ Bethe lattice. The inset shows one standard deviation of $C(t)$ from its average over $100$ realizations. Panel (e) is an abstract generalization to higher $z$ and compares the late-time decay of Eq.~\eqref{eq:effective_X} towards the equilibrium value for different $z$. Best linear fits on a log-log scale (orange) indicate the behavior $t^{-\alpha}$ with $\alpha=1.48,2.99,3.08$ for $z=2,3,4$, respectively.}
    \label{fig:higher_z_fig}
\end{figure*}

\subsection{Hopping on the Bethe lattice}

The central idea of this section is that the combinatorics of reducible correlators of $z$ HI terms finds a natural realization as walks on a Bethe lattice of coordination number $z$ with the $z=2$ case being the 1D walks studied in the previous section. With respect to some fixed origin, we label positions on the Bethe lattice by irreducible words $c$ of $z$ symbols; say $0,1,\dots, z-1$. We denote a single-particle configuration with a particle living on site $c$ by $|c)$. The $z=3$ Bethe lattice is sketched in Fig.~\ref{fig:higher_z_fig}(b). The origin is the empty word $\emptyset$, and the points at ``radius" $1$ (the first dashed ring) are the words $0,1,2$ and so on.

We follow the formulation of \cite{mahan_energy_2001} which studies the tight-binding model on the Bethe lattice by viewing the model as hopping between certain symmetric states that live on a ring at radius $d$ from the origin. These states are equal amplitude and phase superpositions of a particle sitting at all lattice sites at radius $d$:
\begin{equation}
    \ket{d} = P_d^{-1/2} \sum_c |c),
\end{equation}
where the sum runs over radius $d$ words and $P_d$ is the number of such points:
\begin{equation}
    P_d = \begin{cases}
        1\quad &d = 0 \\
        z (z-1)^{d-1} \quad&d \geq 1.
    \end{cases}
\end{equation}
As an example, the first few such states for $z=3$ are:
\begin{align*}
    &\ket{0} = |\emptyset) \\
    &\ket{1} = \big[ |0) + |1) + |2) \big]  / \sqrt{3} \\
    &\ket{2} = \big[ |01) + |02) + |10) + |12) + |20) + |21)\big] / \sqrt{6}.
\end{align*}
Let $\Delta$ be the NN hopping Hamiltonian on the Bethe lattice. The Hamiltonian acts on the $|c)$ states by uniformly hopping to all nearest neighbor points, e.g. for $z=3$ its action is
\begin{align*}
    &\Delta |\emptyset) = |0) + |1) + |2) ,\\
    &\Delta |0) = |\emptyset) + |01) +|02) ,\\
    &\Delta |01) = |0) + |012) + |010),
\end{align*}
etc. Reference \cite{mahan_energy_2001} points out, however, that in the $\ket{d}$ basis, the hopping matrix $\Delta$ becomes simple:
\begin{align*}
    &\Delta\ket{0} = \sqrt{z} \ket{1} \\
    &\Delta\ket{1} = \sqrt{z} \ket{0} + \sqrt{z-1} \ket{2} \\
    &\Delta\ket{d} = \sqrt{z-1} \big[ \ket{d-1} + \ket{d+1} \big]\quad d\geq 2 \\
\end{align*}
and has almost a one dimensional tight-binding form:
\begin{equation}
\Delta = \begin{pmatrix}
0 & \sqrt{z} & 0 & 0 & \cdots \\
\sqrt{z} & 0 & \sqrt{z-1} & 0 & \cdots \\
0 & \sqrt{z-1} & 0 & \sqrt{z-1} & \cdots \\
0 & 0 & \sqrt{z-1} & 0 & \cdots \\
\vdots & \vdots & \vdots & \vdots & \ddots
\end{pmatrix}.
\end{equation}

\subsection{Density of states}

As with the $z=2$ case studied in Sec.~\ref{sec:z2}, we will calculate the moments
\begin{equation}
    \braket{H^n} = \sum_{b\in [z]^n} \braket{h_b},
\end{equation}
where the set $[z] = \{0,1,2,\dots, z-1\}$ and $h_b$ is again shorthand for a correlator defined by the word $b$. Under the assumption that the irreducible correlators vanish in a large $q$ limit, we obtain the correspondence
\begin{equation}
    \braket{H^n} = \braket{0|\Delta^n|0},
\end{equation}
from which it follows that the many-body averaged density of states is equal to the single-particle one for hopping on the Bethe lattice \cite{economou_greens_2006,Giacometti_1995,eckstein_hopping_2005}
\begin{equation}\label{eq:dos_z}
    \mathcal{D}(\epsilon) = \frac{z}{2\pi} \frac{\sqrt{4(z-1) - \epsilon^2}}{z^2-\epsilon^2},\quad |\epsilon|\leq 2\sqrt{z-1}.
\end{equation}
Indeed Fig.~\ref{fig:higher_z_fig}(c) compares this equation with $z=3$ to the ED numerics for three terms in PBC with $q=12$, the result being consistent with irreducible correlators vanishing in the large $q$ limit.

\subsection{Energy dynamics}

One can also consider the energy ``transport" in this model. We focus on infinite temperature and calculate
\begin{equation}
    C(t) = \braket{h(t)h(0)},
\end{equation}
where $h$ stands for any one of the $h_i$ in the Hamiltonian. In a double series expansion, i.e. Eq.~\eqref{eq:double_series}, the combinatorial problem is again to calculate $\braket{H^nhH^mh}$ in the large $q$ limit. Only the reducible correlators survive, and we thus need to count the number of words of the form $aibi$ with $a\in [z]^n$ and $b\in[z]^m$, for some fixed $i$, that are fully reducible. To count these, we require that $a$ reduces to some irreducible word, say $c$, (equivalently a site $c$ on the Bethe lattice) and $ibi$ reduces to the inverse of that word, $c^{-1}$.  Then, we sum over all points $c$. We show the details of this counting problem in Appendix \ref{app:counting} and give the result here.

Let the number of such words be called $F_{nm}$, while $W^d_n$ is the number of fully reducible words starting at the origin and ending at a fixed point at radius $d$ on the lattice. Finally, let $A_d$ be the number of radius $d$ irreducible words which end and begin in the same fixed symbol $i$. Then,
\begin{multline}\label{eq:F_nm}
    F_{nm} = W^0_n W^0_m \\
    + \sum_{d\geq 1}(A_{d} W^{d-2}_n + 2A_{d+1} W^d_n + A_{d+2} W^{d+2}_n)W^d_m,
\end{multline}
where we define $W^{-1}_n=W^1_n$. For both conceptual and practical reasons we would like to understand what ``physical" process this corresponds to for the particle hopping on the Bethe lattice. In terms of the hopping matrix $\Delta$,
\begin{equation}
    W_n^d = \frac{\braket{d|\Delta^n|0}}{\sqrt{P_d}}
\end{equation}
since $\braket{d|\Delta^n|0}$ counts the number of walks that end at distance $d$ up to a normalization. We can identify a non-local operator $X$ such that \footnote{We were not able to identify an operator $V$ such that $C(t) = \braket{0|V(t)V(0)|0}$ as we were in the $z=2$ case, the main issue being that on the Bethe lattice ($z>2$), points are not labeled by $d$ anymore.}
\begin{equation}\label{eq:effective_X}
    C(t) = \braket{0|X(t)|0}.
\end{equation}
This operator can be read off from Eq.~\eqref{eq:F_nm} and is best understood in matrix form. In the $\ket{d}$ basis, we have
\begin{equation}
X = \begin{pmatrix}
1 & 0 & \frac{A_2}{\sqrt{P_0P_2}} & 0 & 0 & \cdots \\
0 & \frac{1}{z} & 0 & \frac{A_3}{\sqrt{P_1P_3}} & 0 & \cdots \\
\frac{A_2}{\sqrt{P_0P_2}} & 0 & \frac{2A_3}{P_2} & 0 & \frac{A_4}{\sqrt{P_2P_4}} & \cdots \\
0 & \frac{A_3}{\sqrt{P_1P_3}} & 0 & \frac{2A_4}{P_3} & 0 & \cdots \\
0 & 0 & \frac{A_4}{\sqrt{P_2P_4}} & 0 & \frac{2A_5}{P_4} & \cdots \\
\vdots & \vdots & \vdots & \vdots & \vdots & \ddots
\end{pmatrix},
\end{equation}
and $C(t)$ can then be understood as the following experiment: a particle begins at the origin at time $-t$. Then, at time $0$, the system is kicked with the operator $X$ everywhere on the lattice; then $C(t)$ is the amplitude for the particle to return to the origin at time $t$.

Since momentum is no longer a good quantum number, an analytical calculation becomes more difficult.  We can, however, run the hopping dynamics numerically on a large but finite Bethe lattice of radius $R$. Introducing a finite-size cutoff means the dynamics will deviate from the infinite lattice case after a certain time $t^* = (2\sqrt{z-1})^{-1} R$, which is the timescale for the particle to propagate ballistically to the edge. In Fig.~\ref{fig:higher_z_fig}(c) we compute $C(t)=\braket{0|X(t)|0}$ for the $z=3$ Bethe lattice for $t < t^*$ and compare it to the numerical solution for three random matrix terms arranged as in Fig.~\ref{fig:higher_z_fig}(a). Here, again a single realization is shown, and we found it necessary to show a very small $q=6$ to observe noticeable error. For $q=10$, the curves were essentially indistinguishable. In the inset, we also show one standard deviation from the average as a function of time, which is small (roughly $1.5 \% $ of the average) and flat on the plotted timescale.

In Fig.~\ref{fig:higher_z_fig}(e) we study the dynamics on the Bethe lattice, independently of which model maps to this picture. We observe that
\begin{equation}
    C(t) \rightarrow \frac{1}{z} + O(t^{-\alpha_z}),
\end{equation}
as $t\rightarrow\infty$. We can see numerically that $\alpha_z\approx 3$ for $z\geq 3$, which can be explained as follows. In \cite{bellitti_2019} it is shown that for any $H$ whose density of states displays a  square-root dependence on energy at the edge of the energy spectrum, the late time behavior of $C(t)$ (specialized to our situation) will be
\begin{equation}\label{eq:surmise}
    C(t) \rightarrow C(\infty) + c_z |\langle e^{-iHt}\rangle |^2
\end{equation}
as $t \rightarrow \infty$ for some constant $c_z$ that depends on $z$. Furthermore, this square-root edge will lead to the asymptotic behavior $ |\langle e^{-iHt}\rangle |^2 \rightarrow \propto t^{-3} $.  Indeed, for $z\geq 3$, the density of states, Eq.~\eqref{eq:dos_z}, has a square-root edge, which can be seen by setting $\epsilon = 2\sqrt{z-1}(1-\delta)$ for a small $\delta>0$: the leading-order term is proportional to $\sqrt{\delta}$. This explains the exponent of $\alpha=3$. The $z=2$ case is rather special, with the slow decay $t^{-3/2}$ of the two-point function, in part owing to the unique inverse square root band edge in the density of states Eq.~\eqref{eq:arcsine}. 

In this totally non-commuting scenario, by writing $H = h + \tilde{h}$, where $\tilde{h}$ stands for all other $z-1$ terms in $H$, we can conjecture that, physically, $z\geq 3$ decays faster than $z=2$ because in the former case, the ``bath" Hamiltonian $\tilde{h}$ is chaotic, while for $z=2$ it is not.

\section{Energy transport in the extended chain}\label{sec:chain}

\begin{figure*}[t]
    \centering
    \setlength{\abovecaptionskip}{-5pt}
    \includegraphics[width=1.5\columnwidth]{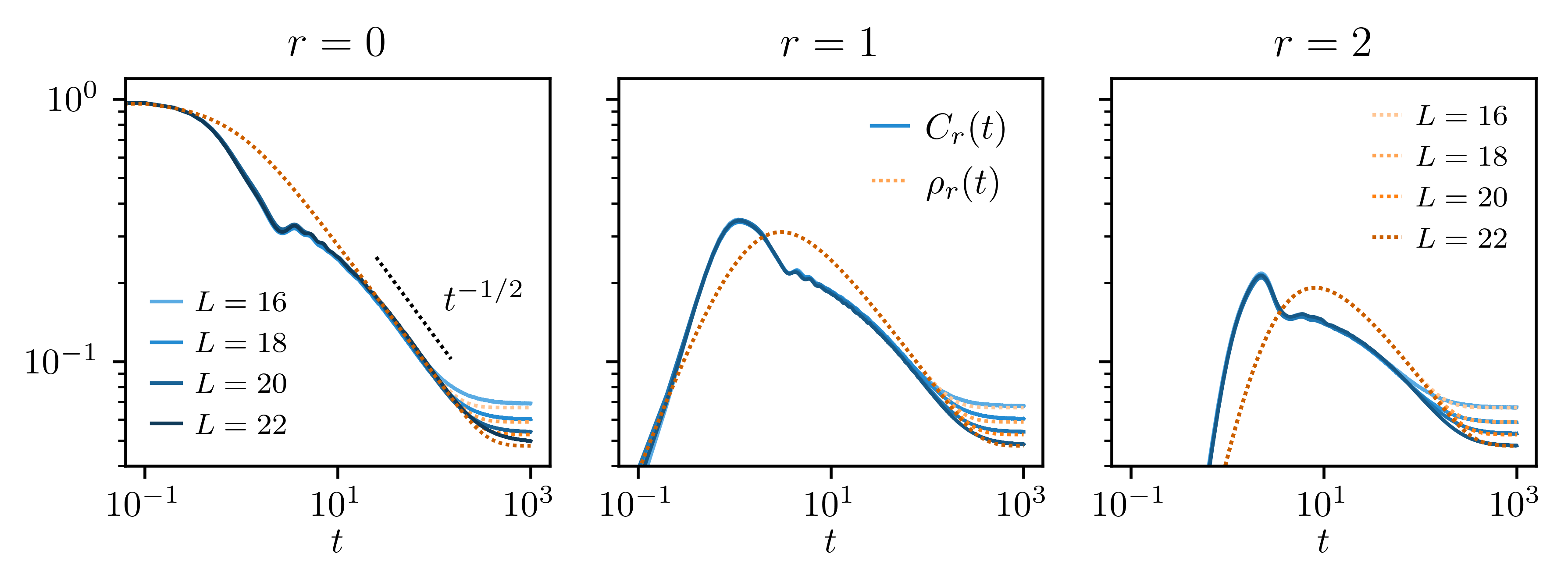}
    \caption{Numerical calculation of energy transport via average two-point functions $C_r(t)$ for a $q=2$ HI RMT chain shown in blue. The results are compared against the solution to the finite-size lattice diffusion equation $\rho_r(t)$, shown in dashed orange for the corresponding values of $L$, with diffusivity $D=0.4$. All data correspond to averages over $1024$ RMT realizations, except $L=22$ and $r=1,2$ correspond to an average over $384$ realizations.}
    \label{fig:transport}
\end{figure*}

So far we have discussed the energy dynamics of some few-body scenarios of Hamiltonians with HI local terms. As a first step towards treatment of the extended system, we study the energy transport in a $q=2$ HI RMT chain numerically. In particular, we would like to confirm the hypothesis that the long-time dynamics of the local energy density should be describable by a simple diffusion equation due to the presence of only energy conservation. We find good evidence of diffusion in the autocorrelator $C_0(t)$ of the energy density, but find that the spatially distant two-point correlators have not yet converged to the solution of the diffusion equation at the considered system sizes. This suggests that for this model, more sophisticated numerical methods will be needed to demonstrate energy diffusion conclusively. 

The observation of emergent hydrodynamics directly from the exact dynamics of two-point functions is well known to be numerically difficult since it requires long times and large systems. For such calculations, we are computationally limited to $q=2$, although small $q$ is of course also of physical interest. To help mitigate finite-size effects, we consider exciting the leftmost bond of the chain in OBC, so that the perturbation has the entire length of the chain into which it can diffuse. In this section, we write
\begin{equation}
    C_r(t) = \braket{h_r(t)h_0(0)}
\end{equation}
to describe the energy response at space-time point $(r,t)$ following the excitation at the boundary.

Another step we take to mitigate finite-size effects is to compare the exact dynamics to the diffusion equation defined on the same discrete and finite-size lattice. We define $\rho_r(t)$ to be the solution of the diffusion equation and directly identify $\rho_r(t)$ with $C_r(t)$ (semi-classically the amount of energy density present on bond $r$). We also write $J_r(t)$ for the current density leaving site $r$ to the right. The diffusion equation combines the continuity equation with Fick's law (diffusion constant $D$) which read
\begin{gather}
    \dot \rho_r(t) = J_{r-1}(t) - J_r(t) \\
    J_r(t) = - D \big[ \rho_{r+1}(t) - \rho_{r}(t) \big].\label{eq:fick}
\end{gather}
Since we want to describe OBC, we require no current entering from the left nor leaving from the right:
\begin{equation}
    J_{-1}(t)=J_{L-2}(t)= 0.
\end{equation}
We emphasize that in the diffusion equation, $r$ refers to a site whereas in the chain $r$ refers to a bond. So in an OBC chain of length $L$, which has $L-1$ bonds, the rightmost site is labeled $r=L-2$. Since the diffusion equation is linear in the density, it is easily solved numerically by treating the equation as an imaginary time Schrodinger equation and applying the propagator to an initial configuration; in our case the initial condition being $\rho_0(0) = 1$ and $\rho_r(0) = 0$ for $r\geq 1$.

For the infinite-temperature spin-chain numerics, we make use of quantum typicality:
\begin{equation}
    \braket{\psi|A|\psi} = \tr[A]+ O\bigg(\frac{\tr[A^\dagger A]^{1/2}}{q^{L/2}} \bigg)
\end{equation}
as $L\rightarrow \infty$ for a single realization of a Haar random state $\ket{\psi}$ of $L$ spins. One can derive this formula using the Weingarten calculus, but the idea of using a random vector to estimate traces goes all the way back to \cite{skilling_maximum_1989}. With this method, one simply needs to calculate the two vectors $h_r e^{-iHt}\ket{\psi}$ and $e^{-iHt}h_0\ket{\psi}$, so that $C_r(t)$ is given by their overlap. While one or only a handful of realizations usually suffice in this context, since we must average over many RMT realizations anyways, we also average over a new Haar random state for each run, which can only improve the accuracy. We evolve the states using Krylov time evolution with a basis dimension of 12 \cite{Park1986}.

In Fig.~\ref{fig:transport} we plot $C_r(t)$ for $r=0,1,2$ on a log-log scale, and compare the dynamics to the numerical solution of the diffusion equation. We set the diffusion constant to $D=0.4$ by eye so that the autocorrelator $C_0(t)$ matches $\rho_0(t)$ for an extended period of time. We can see good convergence of the autocorrelator to the diffusion equation with an $L$-independent diffusion constant. However, for $r=1,2$, we see that, with the choice of diffusion constant set by the autocorrelator, these two-point functions are slower to converge to the solution. We cannot, therefore establish that the local HI RMT chain exhibits diffusive energy transport at the studied system sizes, but we expect a sufficiently large chain to display energy diffusion.

\section{Spectra and Free probability}\label{sec:free}

\begin{figure}[b]
    \centering
    \includegraphics[width=0.95\columnwidth]{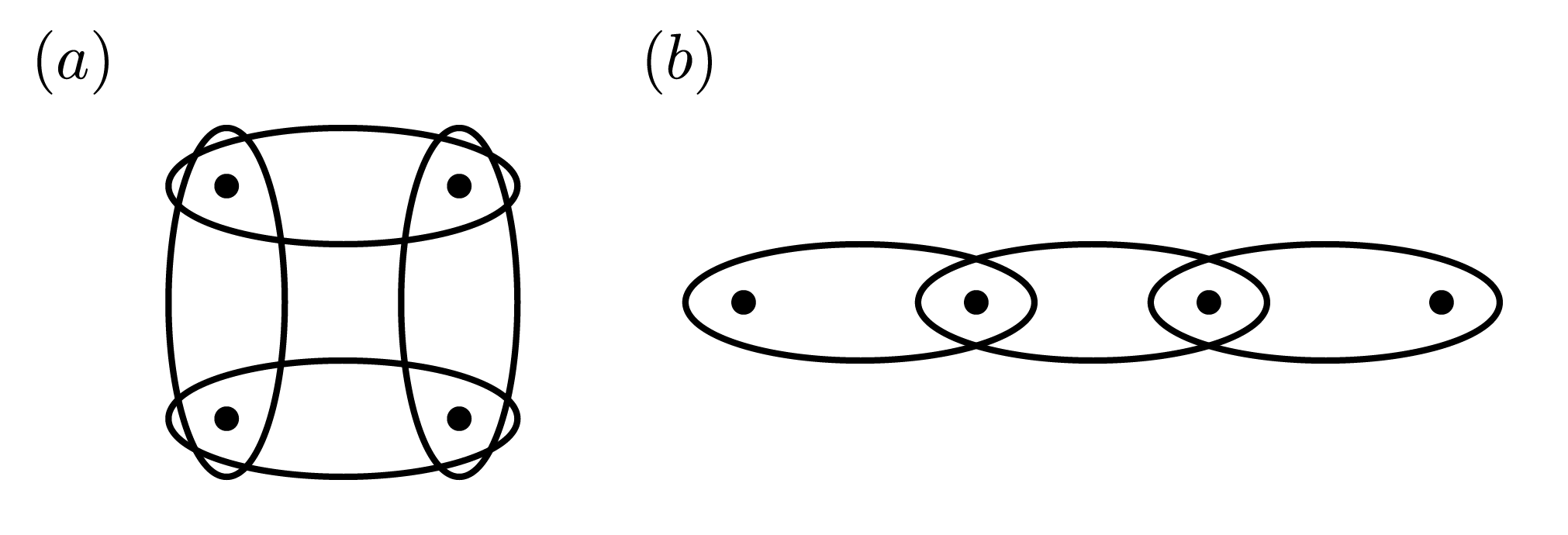}
    \caption{Sketches of two small local HI random matrix Hamiltonians where a free convolution can be readily applied to compute the density of states. Those are calculated numerically and shown, for (a) and (b), in Fig.~\ref{fig:spectra} (a) and (b), respectively. }
    \label{fig:few_body_skecthes}
\end{figure}

\begin{figure*}[t]
    \centering
    \includegraphics[width=2.05\columnwidth]{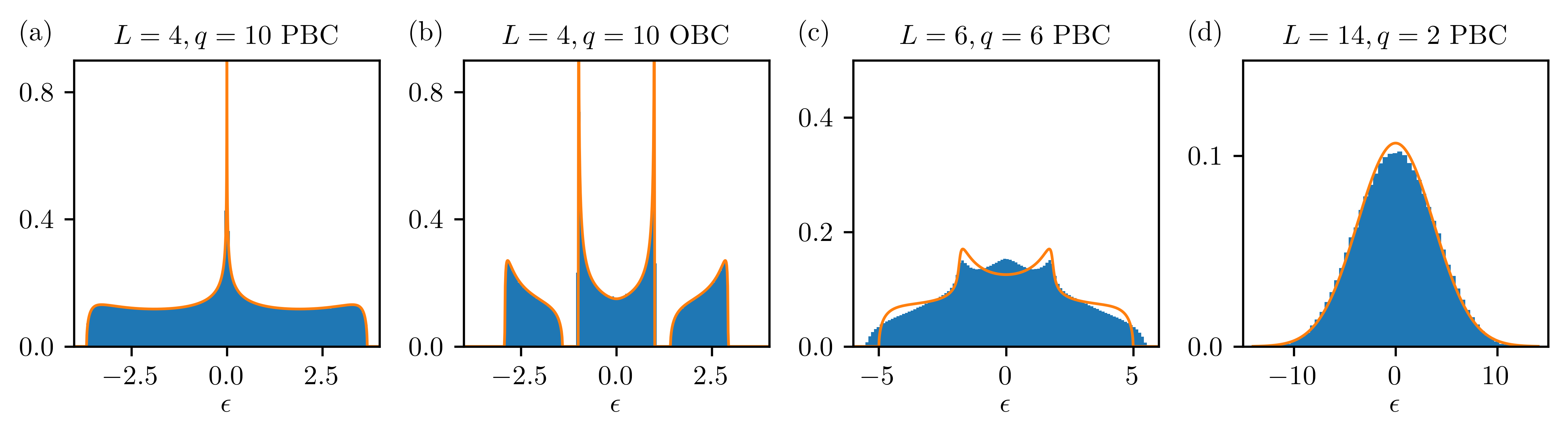}
    \caption{Density of states calculated from ED for single realizations of the HI chain, Eq.~\eqref{eq:full_model}. In panel (a), the orange curve is the a numerical free convolution of two binomial distributions $ \mathcal{B}_2\boxplus  \mathcal{B}_2$ [Eq.~\eqref{eq:binomial_2}]. The curve in panel (b) is a numerical free convolution of a Bernoulli with a binomial,  $\mathcal{B}_1\boxplus \mathcal{B}_2$ and in panel (c) the orange curve is again a free convolution $\mathcal{B}_3\boxplus \mathcal{B}_3$ [Eq.~\eqref{eq:binomial_3}]. Panel (d) is the opposite limit: large $L$ and small $q$. The orange curve is a normal distribution having variance $\sigma^2 = L$.}
    \label{fig:spectra}
\end{figure*}

We have so far studied the energy dynamics in various limits of the HI local RMT. A somewhat simpler question regarding the properties of a local Hamiltonian RMT is the equilibrium physics, determined by the density of states. In this section we make more explicit the connection between our previous language of irreducible words vanishing and the notion of asymptotic freeness. We ask if free probability theory can aid in the computation of the density of states for this model, and if the local HI property can be a useful simplification in this context. We consider both few-body and extended limits of the model and in the few body case we go beyond considering only non-commuting terms as we did in the analytical treatment in Secs.~\ref{sec:z2} and~\ref{sec:higher_z}.

Free probability develops notions of classical probability theory, such as independence, convolution, and the central limit theorem, for non-commuting random variables \cite{nica_lectures_2006,mingo_free_2017}. Random matrices are one key example of non-commuting random variables. The analogue of independence is the notion of freeness. Two non-commutative random variables $A,B$ are free (or freely independent) if any alternating centered correlator of $A,B$ is also centered i.e.
\begin{equation}
    \left \langle \prod_{i=1}^{n}(A^{p_i} - \braket{A^{p_i}})(B^{q_i} - \braket{B^{q_i}}) \right \rangle = 0,
\end{equation}
for all $n \geq 1$.
Many collections of random matrices are known to be asymptotically free, that is, the large matrix size limit of the above correlators vanishes. 

A well-studied question in free probability is: given some large random matrices $A$ and $B$ that have probability distributions $a,b$ for their eigenvalues, what is the probability distribution for the eigenvalues of $A+B$? A well-known result says that if the eigenvectors of $A$ and $B$ are in ``generic position" i.e. related by a full Haar random unitary $U$, then $A$ and $B$ are asymptotically free \cite{mingo_free_2017} and the density of states of their sum is given by the free convolution, denoted $a\boxplus b$, where $a$ is density of states of $A$ and $b$ the density of states for $B$ in the large matrix size limit. The free convolution in some special cases can be computed explicitly, but in general must be done numerically. 

For a pair of overlapping HI terms, we showed in Sec.~\ref{sec:z2} that the alternating correlators of those two terms vanished in the large $q$ limit. A simple consequence of this fact is that the terms are asymptotically free. This follows since
\begin{equation}
    h_i^n - \braket{h_i^n} = \begin{cases}
        h_i \quad n \ \text{odd}\\
        0 \quad n \ \text{even}
    \end{cases} .
\end{equation}
That the two terms are asymptotically free is consistent with our earlier results, as the free convolution of two symmetric Bernoulli distributions is the arcsine distribution \cite{nica_lectures_2006}. In fact, considering $z$ mutually non-commuting HI terms (Sec.~\ref{sec:higher_z}), the statement that irreducible words vanish is also equivalent to the statement that those terms are asymptotically free by an extension of the above argument for two terms.

Once true locality of interaction is introduced into an RMT, however, these tools do not directly apply. While the Hamiltonian will be of the form $A+B+\cdots $, the individual terms are generally no longer free, since their eigenvectors are no longer in generic position. Rather, since most of the terms commute, they actually share eigenvectors, and, if they are drawn independently, then they behave as classically independent random variables. A few works have begun to address this question of the density of states of a local random matrix model in the large $q$ limit, and in particular how to address the mixture of free and classical independence \cite{collins_spectrum_2023,morampudi_many-body_2019,speicher_mixtures_2016}. In the mathematical literature the topic goes by the name $\Lambda-$freeness or $\varepsilon-$freeness \cite{mlotkowski2004lambda,cebron2024graphon}.  In particular, in \cite{collins_spectrum_2023}, their ``XY-model" is identical in structure to our local nearest-neighbor chain, but these works are still inconclusive insofar as producing a concrete example of the density of states of a local random matrix theory.

On the other hand, based on intuition from a non-interacting system of $L$ spins or numerical exact diagonalization of interacting spin chains, one expects that if first the $L\rightarrow\infty$ limit is taken, then the density of states should approach a Gaussian distribution. In fact, a Gaussian density of states was proven for local translation-invariant random $q=2$ spin chains \cite{keating_spectra_2015} and for disordered $q=2$ ``spin-glasses" on graphs with bounded degree in \cite{erdos_phase_2014}. In Appendix~\ref{app:gauss} we also give a heuristic argument for the 1D chain having Gaussian moments, which is independent of the microscopic details and in particular independent of $q$. For the local HI random matrix, one can see in Fig.~\ref{fig:spectra}(d) that a modest $L=14$ chain with small $q=2$ has an approximately Gaussian density of states of variance $\sigma^2=L$. It is still an interesting question as to precisely how this non-interacting character of the density of states emerges as $L$ grows for a particular model, and in particular if the infinite $q$ limit is taken first, as in the free probability literature.

In the remainder of this section, we ask if a naive free convolution of different terms in various small local HI Hamiltonians can produce the correct density of states, empirically. We study short chains, since we expect a long chain to have a Gaussian density of states and therefore these intermediate cases are more interesting. For the free convolution, in most cases this must be done numerically and in Appendix~\ref{app:conv} we outline an efficient numerical method, known as the subordination iteration method \cite{speicher_free_2019}, for obtaining the density of states of a sum of two free variables. We employ this method for studying several systems. 

One might naively attempt to split the nearest neighbor chain into $H=H_\text{even} + H_{\text{odd}}$, where the two terms contain the sum of all even and odd bonds respectively. This division is chosen since all terms in $H_\text{even}$ are mutually commuting, and the same for $H_\text{odd}$. For example, we can take an $L=4$ chain in PBC, which has $4$ bonds, see Fig.~\ref{fig:few_body_skecthes}(a). We can try to split $H$ as
\begin{equation}
    H = h_1 + h_2 +h_3 + h_4 = (h_1 + h_3) + (h_2 + h_4).
\end{equation}
Here, the matrices inside the parentheses both have the distribution
\begin{equation}\label{eq:binomial_2}
    \mathcal{B}_2(\epsilon) = \frac{1}{4} \big[ \delta(\epsilon+2)  + 2 \delta(\epsilon) + \delta(\epsilon-2)\big],
\end{equation}
where $\mathcal{B}_2$ stands for binomial and is the classical convolution of two smaller binomial or Bernoulli distributions
\begin{equation}
    \mathcal{B}_1(\epsilon) = \frac{1}{2}[\delta(\epsilon+1)+\delta(\epsilon-1)].
\end{equation}
One may wonder if the eigenvectors of $(h_1+h_3)$ and $(h_2+h_4)$ are in ``sufficiently generic position" such that the total density of states is given by a free convolution $ \mathcal{B}_2\boxplus \mathcal{B}_2$. In Fig.~\ref{fig:spectra}(a) we find strong numerical evidence that the free convolution produces the correct density of states by comparing it to ED. However, what does not produce the correct density of states is to split the Hamiltonian as
\begin{equation}
    H = (h_1 + h_2) + (h_3 + h_4)
\end{equation}
and then to freely convolve two arcsine distributions, i.e. Eq.~\eqref{eq:arcsine}. Instead, this reproduces the density of states Eq.~\eqref{eq:dos_z} for $z=4$, which is known \cite{novak_three_2012}. We leave the resolution of this issue of the order of convolution for future work. 

We also emphasize here, that the local HI assumption is a practically useful simplification because the classical convolution of $n$ HI terms is trivial. Should we have considered a local GUE Hamiltonian, this would have been more involved, as we would need to compute the classical convolution of many semicircular distributions which is not so simple. 

We can also consider the case of four sites but with three terms in OBC, see Fig.~\ref{fig:few_body_skecthes}(b). There, we split
\begin{equation}
    H = (h_1 + h_3) + h_2
\end{equation}
and so we calculate $ \mathcal{B}_2 \boxplus  \mathcal{B}_1$. Fig.~\ref{fig:spectra}(b) shows this also appears to be the correct distribution. The fact that the energy spectrum has gaps is very interesting. Assuming that irreducible words vanish even in this setting, this suggests the model maps to a particle hopping on a lattice hosting multiple sub-lattices.

The naive splitting into $H=H_\text{even} + H_{\text{odd}}$ and assuming a free convolution of those two terms, however, fails in general, e.g. in Fig.~\ref{fig:spectra}(c) for a slightly larger chain of $L=6$. Here we calculated $ \mathcal{B}_3\boxplus \mathcal{B}_3$ where
\begin{equation}\label{eq:binomial_3}
     \mathcal{B}_3 = \frac{1}{8}\big[\delta(\epsilon+3)+3\delta(\epsilon+1) + 3\delta(\epsilon-1)+\delta(\epsilon-3)\big].
\end{equation}
This is consistent with the mathematical literature on mixtures of free and classical independence \cite{speicher_mixtures_2016} where it is first noted that the situation cannot be split into two groups. These chain configurations with more than two terms serve as non-trivial examples of asymptotic $\varepsilon$-independence. As mentioned before, however, the development of this framework is in its early stages and there do not seem to be any results explicitly computing the density of states. It may be possible to develop a numerical $\epsilon$-free convolution similar to the subordination iteration method. 

\section{Conclusion and Outlook}\label{sec:conc}

In this paper we have studied the evolution of the local energy density under the dynamics of a local random matrix Hamiltonian whose bond operators square to one. This simplification allowed for the analytical calculation of the density of states, a two-point function, and an OTOC for the two-term and large $q$ scenario (Sec.~\ref{sec:z2}). We found a similar mapping of a generalized model with $z$ non-commuting terms (Sec.~\ref{sec:higher_z}) onto walks on the Bethe lattice and we calculated energy-energy correlators, finding that the two point function always decays as $t^{-3}$ for any $z\geq 3$. We also examined energy transport in the small $q$ limit and with increasing $L$, finding evidence of a slow convergence towards diffusion, but the data are insufficient to conclude that energy transport in this model falls within the universality class of diffusion. We then  moved on to study the simpler question of the density of states in Sec.~\ref{sec:free}. We discussed connections to free probability theory, and showed numerically that in some few-body cases, a classical convolution of the Hamiltonian terms followed by a free convolution can yield the correct density of states despite the technical issue of some terms commuting and some not, but that this method fails in general. We conclude with a few comments and ideas moving forward.

Looking towards the full 1D local HI RMT model, the fact that only reducible words survive the large $q$ limit will be crucial to its solution. That only the fully reducible words survive in general, even when some terms commute and some do not, follows from recent work in $\varepsilon$-free probability theory \cite{collins_spectrum_2023,charlesworth_matrix_2021} as applied to our simplified situation where $h_i^2=1$. To confirm this on a numerical level, in Appendix.~\ref{app:irr} we look explicitly at $L=4$ and $L=6$ PBC HI RMT chains and find that only the reducible words contribute to the moments as $q\rightarrow\infty$.

With this in mind, a mapping to a particle hopping on more general graphs may be obtained. It is interesting that the density of states for the $L=4$ PBC chain [Fig.~\ref{fig:spectra}(a)] has a peak in the middle of the band which is reminiscent of the van-Hove singularity for single-particle hopping in 2D \cite{economou_greens_2006}, while the $L=6$ PBC chain [Fig.~\ref{fig:spectra}(c)] also has the qualitative shape of the density of states for hopping in 3D, i.e. a relatively flat central region. We can speculate that the density of states for $L$ terms in PBC could be derived by considering hopping on some kind of $L/2$ dimensional hypercube-like lattice, where the trajectories on this lattice compute the combinatorics of reducible words. This picture is consistent with expectations in the sense that as we take the dimensionality of the hopping to infinity, the density of states approaches a Gaussian.

\begin{figure}
    \centering
    \setlength{\abovecaptionskip}{-5pt}
    \includegraphics[width=\columnwidth]{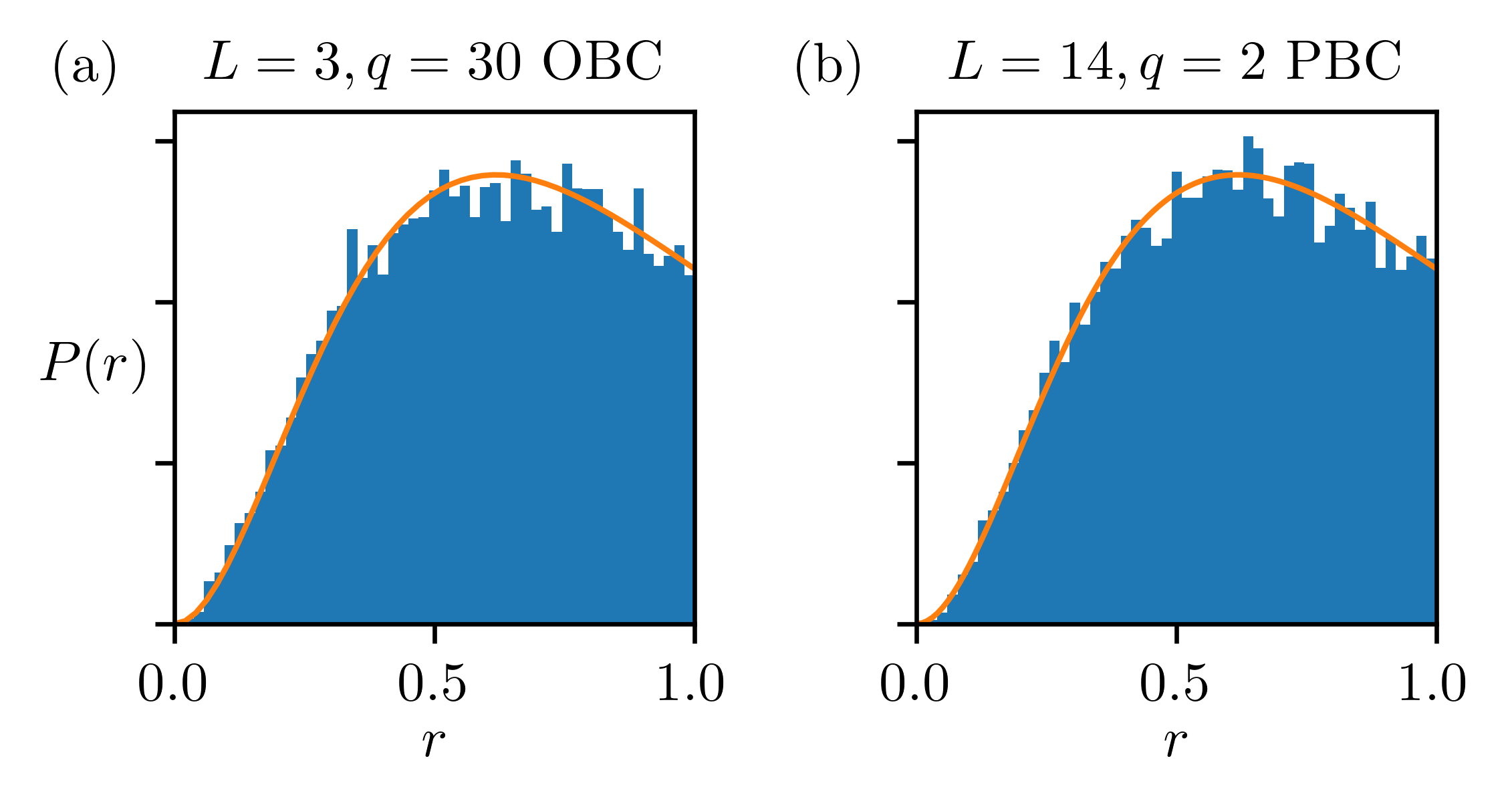}
    \caption{Statistics $P(r)$ of the level spacing ratio $r$ [Eq.~\eqref{eq:ratio}] for the local HI chain. A single realization of the two term case at large $q$ is shown in panel (a) and the extended chain at small $q$ in panel (b). The orange curves are the standard surmise distribution for the GUE.}
    \label{fig:level_spacing}
\end{figure}

We have not touched on quantum chaos. Given that the HI terms square to one, and that the single trace quantities we have calculated admit a description in terms of a single free particle, which is certainly not a quantum chaotic system, one may wonder about the chaoticity of the model. In Fig.~\ref{fig:level_spacing}, we show, however, that the level spacing statistics are that of a GUE in both the few-body and extended limits of the model. In this figure we computed $50$-bin histograms of the level-spacing ratio
\begin{equation}\label{eq:ratio}
    r_n = \frac{\min(s_{n+1},s_n)}{\max(s_{n+1},s_n)}, \quad s_n = E_{n+1} - E_n,
\end{equation}
and compared against the standard expression for $P(r)$ derived in \cite{atas_distribution_2013}. To further diagnose chaos, we need to calculate multi-trace quantities such as two-point and higher-point spectral form factors (SFFs). Is there a mapping to the hopping picture for two-trace quantities like the two-point SFF, at least for the two-term model? Such a calculation should involve counting words that are not fully reducible, and will be the subject of future work.

\acknowledgments

The authors are thankful for fruitful discussions with John Chalker, Weitao Chen, Anatoly Dymarsky, Joseph Eix, Rebecca Flint, Siddharth Jindal, Alexey Khudorozhkov, Aleksei Khindanov, Toma\v{z} Prosen, Rustem Sharipov, Roland Speicher, Anastasiia Tiutiakina, Iris Ul\v{c}akar, and Marko \v{Z}nidari\v{c}. We thank Karen TerHorst for assistance with the graphics. K.P. and T.I. acknowledge support from the National Science Foundation under Grant No. DMR-2143635. J.K. acknowledges support from the U.S. Department of Energy under Grant No. DE-SC0023692. N.P. acknowledges support from the Natural Sciences and Engineering Research Council of Canada. J.R. acknowledges financial support from the Royal Society through the University Research Fellowship No. 201101. We also acknowledge the hospitality of the Boulder Summer School 2023: ``Non-Equilibrium Quantum Dynamics" with support from the National Science Foundation Grants No. DMR-1560837 and DMR-2328793, where this project was conceived.

\begin{appendix}

\section{Counting on the Bethe lattice}\label{app:counting}

In this section, we detail the counting of the number of words $aibi$ with $a\in [z]^n$ and $b\in[z]^m$, for some fixed $i$, that are fully reducible. To count these, we require that $ibi$ reduces to some irreducible word, say $c$, (equivalently site on the Bethe lattice) and $a$ reduces to the inverse of that word, $c^{-1}$, i.e. $c$ read backwards. Then, we sum over all points $c$. As an equation, we have
\begin{equation}\label{eq:fnm}
    F_{nm} =  \sum_c \sum_a \delta(\text{red}(a) - c^{-1}) \sum_b \delta(\text{red}(ibi)-c).
\end{equation}
Here, $\text{red}(\cdot)$ means apply the algebraic relations $i^2=\emptyset$ to all symbols in the word until no further reduction is possible. The function $\delta(\text{red}(a) - c^{-1})$ is $1$ if $a$ reduces to $c^{-1}$ and $0$ otherwise, and the sum over $c$ runs over all fully irreducible words (equivalently sites on the Bethe lattice).

The next step is to split the irreducible words $c$ into distinct cases, where the endpoints of $c$ are or are not the fixed and privileged symbol $i$. We can have $c=i\tilde{c}i,c=i\tilde{c}j,c=j\tilde{c}i$, or $c=j \tilde{c} k$
where $i\neq j,k$. The second and third cases are equivalent. Let us count the number $A_d$ of lattice points at radius $d$ which end and begin with $i$. Assuming $d\geq3$, this number satisfies the recursion
\begin{equation}
    A_d = (z-1) A_{d-2} + (z-2) A_{d-1},
\end{equation}
since we can start with an irreducible word ending in $i$ and add $ji$ for $j \neq i$, or we can start with a word ending in $i$, remove it (the word now ends in $k$), and add $ji$ where $j\neq k\neq i$. With initial conditions $A_1=1$ and $A_2=0$, the recursion is solved by
\begin{equation}
    A_d = z^{-1} \bigg((z-1)^{d-1} + (z-1) (-1)^{d-1} \bigg),
\end{equation}
which is valid for $d\geq1$. At the same time, the number of words $c=ixj$ for $j\neq i$ is the total number of points starting with $i$, minus those both starting and ending in $i$:
\begin{multline}
    B_d = (z-1)^{d-1} - A_d \\
    = z^{-1}\bigg( (z-1)^d + (z-1)(-1)^d\bigg).
\end{multline}
Finally, there are the words starting and ending in something other than $i$, say they are $C_d$ in number. We must have that
\begin{equation}
    A_d + 2B_d + C_d = P_d
\end{equation}
so we can solve for
\begin{equation}
    C_d = z^{-1}\bigg( (z-1)^{d+1} + (z-1)(-1)^{d+1}\bigg),
\end{equation}
which is again valid for $d\geq 1$.

Finally, let $W^d_n$ be the number of fully reducible words starting at the origin and ending at a fixed point at radius $d$ on the lattice. Then by counting the four cases for $c$ in Eq.~\eqref{eq:fnm} and using $A_d = B_{d-1}$ and $C_d=B_{d+1}$, we obtain
\begin{multline}
    F_{nm} = W^0_n W^0_m \\
    + \sum_{d\geq 1}(A_{d} W^{d-2}_n + 2A_{d+1} W^d_n + A_{d+2} W^{d+2}_n)W^d_m,
\end{multline}
where we have defined $W^{-1}_n\equiv W^1_n$.

\section{Numerical free convolution}\label{app:conv}

Here we briefly review a very simple and powerful numerical method for obtaining the Green's function (a.k.a. resolvent or Cauchy transform) of a sum of two free variables. This method was brought to the authors' attention in the lecture notes \cite{speicher_free_2019}.

Consider two free variables with distributions $\mu$ and $\nu$. They have Green's functions  $G_\mu(u)$ and $G_\nu(u)$. We would like to obtain $G_{\mu\boxplus\nu}(u)$, i.e. the Green's function for the free convolution $\mu\boxplus\nu$.  The idea is to introduce ``subordination" functions $\omega_\mu(u)$ and $\omega_\nu(u)$, defined such that
\begin{equation}
    G_{\mu\boxplus\nu}(u) = G_\mu(\omega_\mu(u)) = G_\nu(\omega_\nu(u)).
\end{equation}
By setting $H_\rho(u) =  u + 1/G_\rho(u)$ for $\rho=\mu,\nu$, one shows via Voiculescu's $R$-transform method that
\begin{equation}\label{eq:iterate}
    \omega_\mu(u) = u + H_\nu(u+H_\mu(\omega_\mu(u))).
\end{equation}
For any fixed $u\in \mathbb{C}^+$ (the upper half of the complex plane), this equation is a fixed point equation for the number $\omega_\mu(u)$. If one iterates this equation for an initial guess $\omega_0 \in \mathbb{C}^+$, it will converge. Then, by sweeping through $u$ one can obtain the functional dependence $\omega_\mu(u)$ and thus $G_{\mu\boxplus\nu}(u)$. The density of states is
\begin{equation}
    \mathcal{D}(\epsilon) = -\frac{1}{\pi} \text{Im}~G_{\mu\boxplus\nu}(\epsilon + i0^+),
\end{equation}
showing that one really only needs to sweep through $u$ that are close to the real axis.

To obtain the orange curves in Fig.~\ref{fig:spectra}(a,b,c), we used the initial guess $\omega_0=1+i$. For each fixed $\epsilon$, we set $u=\epsilon+i0^+$ and then iterated Eq.~\ref{eq:iterate} until $|\omega^n_\mu(u)-\omega^{n-1}_\mu(u)|<0^+$ where $n$ refers to the $n$-th iteration. Here, we took $0^+=10^{-7}$. 

\section{Argument for Gaussian moments}\label{app:gauss}

In this section we give a simple informal argument for the full large $L$ chain to have moments matching that of the Gaussian distribution for any independent local terms satisfying $\braket{h_i}=0$ and $\braket{h_i^2}=1$ \footnote{We thank John Chalker for pointing out the beginnings of this argument to us.}. The moments
\begin{equation}
    \braket{H^n} = \sum_{b\in[L]^n} \braket{h_b}
\end{equation}
are a sum of all possible words. We would like to consider large $L$, and we should consider words which are pairings, since this maximizes the number of free indices (we cannot have an unpaired symbol because $\braket{h_i}=0$). Let $P_n \subset [L]^n$ be the set of pairing and non-colliding words. Here, a pairing word is a word of even length $n$, with $n/2$ free and distinct indices, each of which appears in two positions along the word. A non-colliding word $b$ is a word in which all letters of the word are mutually far enough apart from each other such that all terms in $h_b$ commute.

If $i_1, i_2, i_3, \dots ,i_{n/2}$ are the free indices in the pairing, then we would like to require $|i_k-i_l|\geq 2$ for all $k,l$. Without the non-colliding constraint, there are $L(L-1)(L-2)\cdots (L-n/2+1)$ such configurations. Considering $n\ll L$, this tends to $L^{n/2}$. With the minor constraint $|i_k-i_l|\geq 2$, the number of configurations of this form is smaller. However, the leading order contribution will come from terms where all $i_k$ are spaced far enough apart from each other and will be
\begin{equation}
    L(L-3)(L-6)\cdots (L- 3n/2 + 1),
\end{equation}
which also tends to $L^{n/2}$. The same argument would apply for $|i_k-i_l|\geq \xi$ when $\xi \ll L$. The reason this argument is heuristic is that we need to first take $L\rightarrow\infty$ and then take $n\rightarrow\infty$. The number of pairings is $(n-1)!!$, and so
\begin{equation}
    |P_n| \rightarrow L^{n/2} (n-1)!!.
\end{equation}

\begin{figure}[t]
    \centering
    \includegraphics[width=0.85\columnwidth]{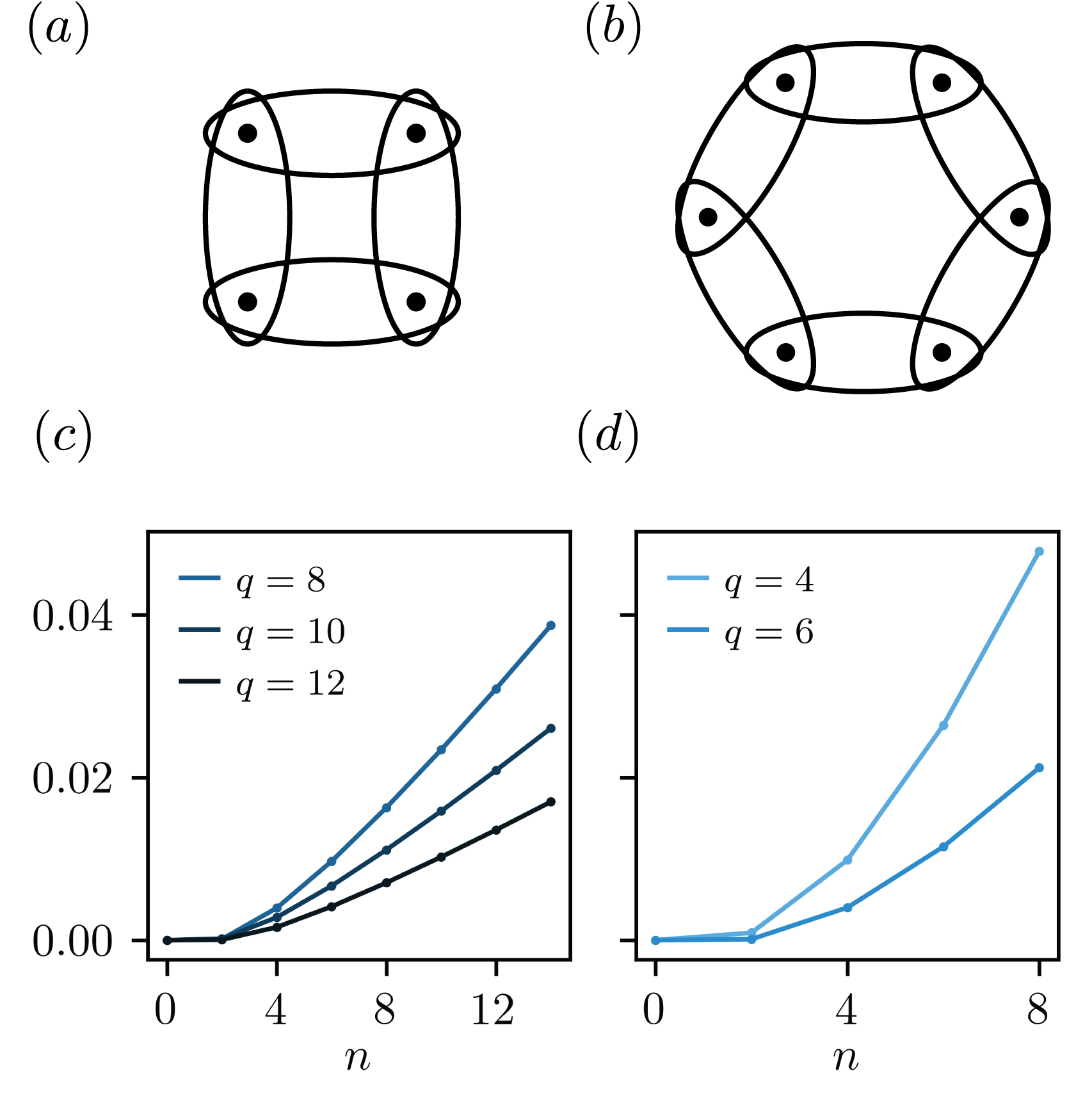}
    \caption{Numerical evidence that only reducible words contribute to the moments as $q \rightarrow \infty$. Panels (a) and (b) are the two RMT models considered, and panels (c) and (d) are the respective \emph{fractional error} between the number $R_n$ of reducible words of length $n$ and the RMT average moments $\braket{H^n}_{100}$, the latter being an empirical average over $100$ realizations. Darker blue is larger $q$; the fractional error systematically decreases with $q$.}
    \label{fig:moments}
\end{figure}

It remains to show that the other types of words are subleading in number. The largest possible contribution from the other words will still be pairings, so it suffices to show that pairing and colliding words are subleading. But, at this point, it is clear that even a single ``collision" of indices in such a word will connect those two indices, making them no longer both free, and the number of such words must then be $O(L^{n/2-1})$ which shows that we still have
\begin{equation}
    \braket{H^n} \rightarrow  L^{n/2} (n-1)!! ,
\end{equation}
which are the even $n$ moments of a Gaussian with variance $\sigma^2 = L$.

\section{Irreducible correlators vanish.}\label{app:irr}

In this appendix, we give some numerical evidence that only irreducible words of the local terms survive the large $q$ limit. Consider the local terms $\{h_j\}_{j=1}^L$ with PBC so we have that $h_ih_j=h_jh_i$ for $|i-j|>1$ (regarding $1\equiv L$). We numerically compute the first few average moments, $\braket{H^n}$, of $L=4$ and $L=6$ chains in PBC and compare them against the number reducible words of length $n$. Since the RMT average moments of $H$ are a sum of all possible words, agreement of the two calculations is good evidence that the irreducible words vanish and only the reducible ones contribute to the moments as $q\rightarrow\infty$.

The number of reducible words, $R_n$, of $L$ terms in PBC having length $n$ are counted symbolically by generating all possible words, then performing two sorting and reduction steps repeatedly until the word can no longer be changed. We assume the word contains an even number of each term, since it will otherwise be trivially irreducible. We then sort the terms (letters) by moving smaller indices to the left, with the constraint that we can only swap adjacent terms $h_i,h_j$ if $i\neq j \pm 1$. The sorted word is then simplified using $h_j^2 = 1$ by removing any pairs of adjacent identical terms. We then apply the same algorithm, but this time sorting the string largest to smallest, and again we remove paired adjacent terms. This is repeated until no further reductions are possible. If this results in an empty string we have found a reducible word. Otherwise we conclude the word is irreducible. 

Fig.~\ref{fig:moments} panels (c) [corresponding to model (a)] and (d) [corresponding to model (b)] show the fractional error
\begin{equation}
    |\braket{H^n}_{100} - R_n|/R_n
\end{equation}
where $\braket{H^n}_{100}$ is an empirical average over $100$ realizations of the RMT. We see that the fractional error is on the order of one percent, and that the error systematically decreases with $q$.

\end{appendix}

\bibliography{references}

\end{document}